\documentclass[aps,twocolumn,prb,showpacs,superscriptaddress,groupedaddress,eqsecnum]{revtex4}  
\usepackage{graphicx}
\usepackage{dcolumn}  
\usepackage{bm}  
\usepackage{amssymb}
\usepackage{amsmath}

\begin{document}
\newcommand{\niceref}[1] {Eq.~(\ref{#1})}
\newcommand{\fullref}[1] {Equation~(\ref{#1})}
\title{Quantum Memory Effects in Disordered Systems and Their Relation to $1/f$ Noise}

\date{\today}

\begin{abstract}
We propose that memory effects in the conductivity of metallic systems can be produced by the same two levels systems that are responsible for the $1/f$ noise. Memory effects are extremely long-lived responses of the conductivity to changes in external parameters such as density or magnetic field.  Using the quantum transport theory, we derive a universal relationship between the memory effect and the $1/f$ noise. Finally, we propose a magnetic memory effect, where the magneto-resistance is sensitive to the history of the applied magnetic field.
\end{abstract}

\author{Yonah Lemonik}
\email{lemonik@phys.columbia.edu}
\author{Igor Aleiner}
\email{aleiner@phys.columbia.edu}
\affiliation{Department of Physics, Columbia University, New York, 10027 USA}
\maketitle

\section{Introduction\label{sec:intro}}

 There are several phenomena in electronic systems that occur on extremely long time scales. 
One well-known example is the $1/f$ noise\cite{DuttaHorn81} where the power spectrum of the conductivity noise shows power law scaling in a range of frequencies from $1\times10^5$Hz to  $1\times10^{-6}$Hz. 

\begin{figure}
\includegraphics[width=\columnwidth]{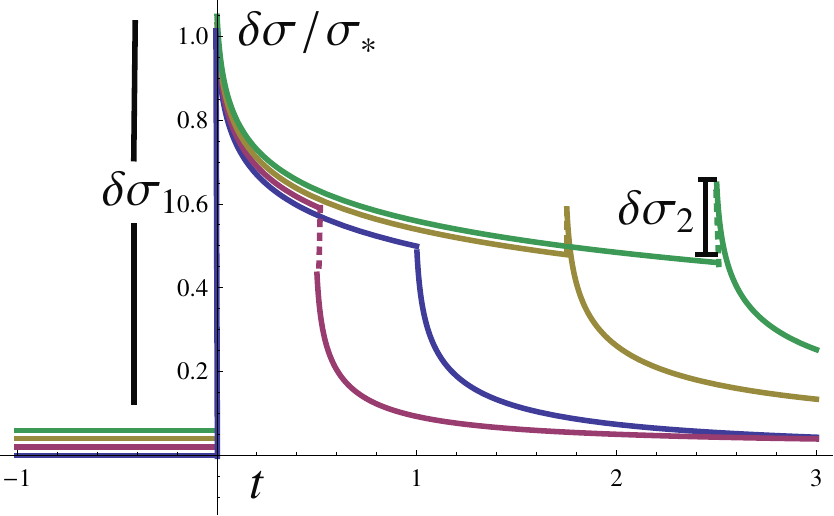}
\caption{\label{fig:PlotCondQuench}Figure showing the reponse of the conductivity to a change in the density $n_e$. The behavior is quaitatively similar for a change in magnetic field. 
The density is changed by $\delta n_e$ at $t=0$ and returned to its original value at $t=t_h$.   
The graph plots conductivity vs. time for several different choices of $t_h$, but the same $\delta n_e$. 
There is a jump in the conductivity $\delta\sigma_1$ when the chemical potential is first changed and a second jump $\delta\sigma_2$ at $t=t_h$. The time scale is in arbitrary units. Figures offset slightly for clarity. The scale $\sigma_*$ is defined in Eq.~(\ref{eq:defSigmaStar}). A positive $\sigma_2$ only appears when $t_h > \sqrt{t_i t_f}$ when 50\% of the TLS are relaxed.
}
\end{figure}

Another such phenomenon is the conductivity memory effect\cite{Grenet12,Grenet03,Martinez97}, where after a sudden change of the electron density the conductivity will jump above its equilibrium value, as illustrated in Fig. \ref{fig:PlotCondQuench}.
The conductivity will relax to its equilibrium value very slowly, without any visible time scale.  Anomalies at the old Fermi level  (see Fig \ref{fig:PlotCondRelax}) may remain detectable up to a day later. 

In the case of $1/f$ noise, it has been proposed\cite{FengLee86,AltshulerSpivak85,ImryBook} that these scales come from two-level systems\cite{Phillips87,Anderson72, Black81} (TLS) with a broad spectrum of tunneling times. The prototypical example of such a TLS is an impurity tunneling between a close pair of host sites. The reaction of the electrons to this motion naturally reproduces the $1/f$ noise.

In this paper we show that this mechanism \emph{by necessity} produces a conductivity memory effect. 
The effect is, in a sense, the inverse of the $1/f$ noise, as it derives from the reaction of the TLS to the mesoscopic fluctuations of the electron density.  As a mesoscopic phenomenon, it is sensitive to magnetic fields and a change in the magnetic field produces qualitatively similar behavior as a change in electron density.
 Moreover we  derive a ``memory magneto-resistance", where the magnetoresistance depends on the history of the magnetic field.

Since the $1/f$ noise and memory effect derive from the same interaction we can derive a ``universal" relationship between the noise and the memory effect, independent of the microscopic details of the TLS. 
 This relationship depends only on the phase coherence length, as measured by the magneto-resistance. 

The plan of the paper is as follows. 
In Section \ref{sec:Results} we give a qualitative discussion of the model and the results. In Section~\ref{sec:Diagrammatics} we give a quantitative derivation of these results using the standard quantum theory of metals. We also analyze the effect of magnetic fields and derive the memory magneto-resistance effect. A derivation of the properties of the TLS is given in Appendix~\ref{sec:TLS}. In Appendix~\ref{sec:Experimental} we discuss an experimental protocol for detecting the memory effect.

\section{Qualitative discussion and results\label{sec:Results}}
The purpose of this section is to review known facts about the $1/f$ noise and make a connection to the proposed memory effect.
\subsection{$1/f$ noise and mesoscopic corrections} 
It has been known for over 50 years that the conductivity noise in metals has strange behavior in the low-frequency limit\cite{DuttaHorn81}. Consider a sample of linear dimension $L$ with a fixed voltage applied such that a mean current $I$ is produced. If the fluctuations of the current around the mean $\delta I(t)$ are measured it is found that,
\begin{equation}
\overline{\delta I(t) \delta I(t')} = I^2 L^{-d}\mathcal{F}(t-t')\label{eq:DefCalF},
\end{equation}
where $\overline{\cdots}$ denotes the time average. The factor of $L^{-d}$ takes into account the central limit theorem so that the function $\mathcal{F}$ does not depend on the sample geometry. The Fourier transform of $\mathcal{F}$ was found to behave as
\begin{equation}
\int \!dt\,\mathcal{F}(t)e^{i\omega t} \sim \frac{1}{|\omega|} \end{equation}
at low frequencies $\omega=2\pi f$.  This behavior persists in some samples from frequencies of a khZ to an inverse day. 
 The basic problem is a mismatch of scales. 
The typical elastic scattering times are of the order of picoseconds. The inelastic scattering (either the dephasing or the energy relaxation time) may exceed the elastic scattering by several orders of magnitude.
 But even these are never larger than a microsecond. How can there be behavior on times of an inverse day? What scale can be the cutoff for the $1/f$ behavior?

A resolution of this problem  has two components. The first component is the two-level system\cite{Phillips87,Anderson72,Black81} (TLS). 
There are many possible microscopic mechanisms that produce appropriate TLSs. 
As our final results should be independent of the microscopic details we will work with a particularly simple model.
This is a heavy but mobile atom with two equilibrium positions $r_1$ and $r_2$. Under the action of inelastic scattering by electrons and phonons the atom can switch its position. 

The probablistic description of the TLS is the following: $P^{eq}_{1,2}$ are the probability for the TLS to be in states $1,2$ as dictated by the Gibbs distribution. 
The motion between these states is characterized by $P(t,r|t's)$, the conditional probability to be in state $r$ at time $t$  provided that it was in state $s$ at time $t'$. A particular TLS is governed by a single relaxation time $\tau_{12}$,

\begin{equation}
P(t,r|t',r) = P^{eq}_r + (1-P^{eq}_r)e^{-|t-t'|/\tau_{12} }.\label{eq:CondPrb}
\end{equation}

The TLS transitions necessarily involve tunneling. 
Therefore the relaxation time $\tau_{12}$ must be of the form,
\begin{equation}
\frac{1}{\tau_{12}} \propto \exp\left(-\frac{|\vec{r}_1-{r}_2|}{a}\right),\end{equation}
where $a$ is a constant on the order of the lattice constant. Assuming that the positions $r_{1,2}$ are homogeneously distributed we find that the probability distribution of the relaxation times is
\begin{equation}
d\tau_{12} \mathcal{P}(\tau_{12}) \sim \frac{d\tau_{12}}{\tau_{12}}\label{eq:TLSDist}.\end{equation}

Averaging \niceref{eq:CondPrb} over TLS with the distribution (\ref{eq:TLSDist}) gives 
\begin{equation}
\int d\tau_{12} \mathcal{P}\left(\tau_{12}\right) e^{-t/\tau_{12}} \propto \frac{\ln\left(t_f/t\right)}{\ln\left(t_f/t_i\right)}=\mathcal{K}(t)\label{eq:DefCalK},
\end{equation}
valid when $t_i < t < t_f$. 
The lower cutoff $t_i$ is given by some microscopic scale and the upper cutoff $t_f$ is larger than $t_i$ by many orders of magnitude in reasonable models.  
The function $\mathcal{K}(t)$ therefore shows the $1/f$ behavior over an extremely large range of scales that is characterstic of $\mathcal{F}(t)$. 
If there were a mechanism that would tranlsate the motion of the TLS into an observable transport coefficient of electrons, we could write $\mathcal{K}(t) \propto \mathcal{F}(t)$ and claim the phenomena explained.

Such a translation is in fact subtle.
 Naively, the conductivity is determined by the Drude formula,
\begin{equation}
\sigma_D  = e^2\nu v^2_F\tau_{tr}\label{eq:Drude},\end{equation}
where  $\nu$ is the density of states, $v_F$ the Fermi velocity and the transport time $\tau_{tr}$ is given by
\begin{equation}
\frac{1}{\tau_{tr}} = v_F N_{imp} s\label{eq:tautr},
\end{equation}
where $N_{imp}$ is the density of impurities and $s$ is the scattering cross-section. 
Given that shifting an impurity does not change its scattering cross-section\cite{LI}, it would seem that the motion of the impurity has no effect on the conductivity at all.

It was realized in Refs.~[\onlinecite{FengLee86,AltshulerSpivak85}] that the theory of meseoscopic conductance fluctuations\cite{LeeStone85,FukuyamaLeeStone87,Altshuler85} resolves this issue. 
To illustrate this resolution let us recall the justification for the Drude equation. 
The Fermi wavelength $\lambda_F$ is much smaller the mean free path between impurities $\ell_{imp}$, so we may consider the electrons as wavepackets following semiclassical trajectories. 
Consider the probability $W_{AB}$ for an electron to propagate from point $A$ to point $B$.
 Because the electrons can scatter off an impurity to any direction there are many paths connecting the two points. 
Quantum mechanically, we assign to each path $i$ the amplitude $\mathcal{A}_i$, sum the amplitudes, and square the result. This gives,
\begin{equation}
W_{AB} = \sum_i|\mathcal{A}_i|^2 +\sum_{i\neq j}\mathcal{A}^*_i\mathcal{A}_j.\label{eq:Prob}\end{equation}
\begin{figure}
\includegraphics[width=\columnwidth]{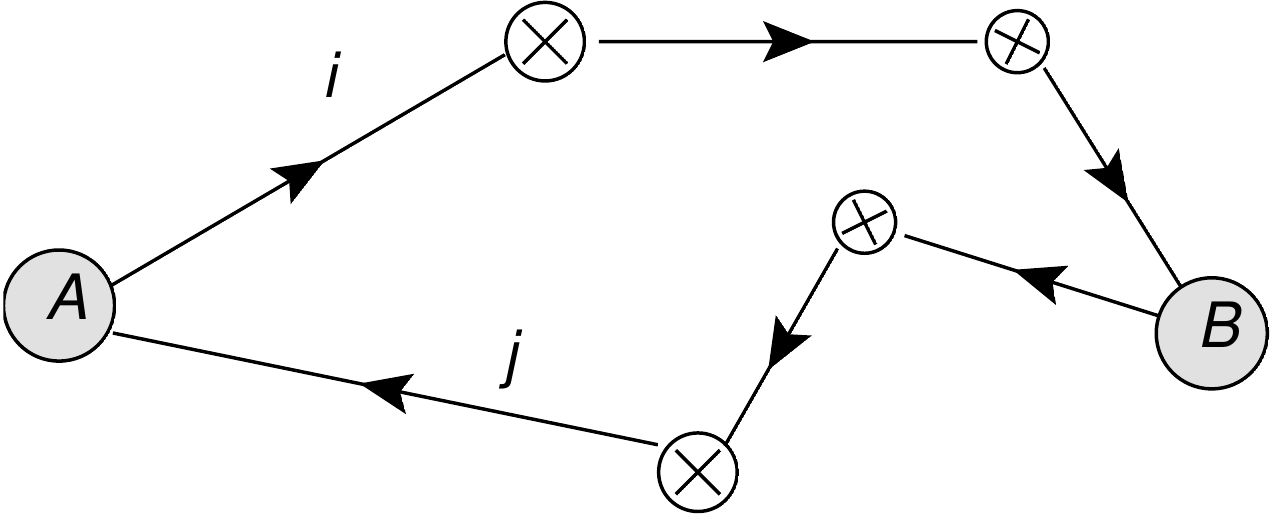}
\caption{\label{fig:Paths1}An illustration of semi-classical paths in the ``interference" contribution to the probabiliy to propagate from point A to point B. The crossed circles represent static impurities and the reversed arrow indicates the complex conjugate of the amplitude.
}
\end{figure}
The first term is a classical sum of probabilities which leads to the diffusion equation and the Drude formula. The second ``interference term", illustrated in Fig. \ref{fig:Paths1}, is neglected in the Drude equation.
 The usual justification is that the interfence depends on the relative phase of two paths,
\begin{equation}
 \phi_{ij} \sim (L_i - L_j)p_F/\hbar, \end{equation}
where $L_i$ is the length of the $i$th trajectory and $p_F$ is the Fermi momentum.  But this phase fluctuates wildly since $p_F L_i \gg \hbar$. Thus one may think, incorrectly, the interference correction is a sum of terms with random signs and may be neglected. 
The remaining terms are purely classical and so any correction to the conductance $G$ would take the form,
\begin{equation}
\delta G \overset{?}{\sim}\frac{1}{N}\sum_{i}\left(|\mathcal{A}_i|^2 +\delta g_i\right)\label{eq:ClassicalG1},
\end{equation}
where $N$ is the number of paths and $\delta g_i$ is a correction to the classical probability. This leads to a variance
\begin{equation*}
\langle \Delta G^2 \rangle \overset{?}{\sim} \langle \delta g^2_i \rangle \frac{1}{N^2} N\propto\frac{1}{N}.\end{equation*}
Thus, according to this logic, the correction to the conductivity decays with $N$. Since $N$ grows with the size of the system, this leads one to think that all corrections must decay with the size of the system.

However, the neglect of the interference term above is careless, since there are pairs of paths whose phases are fixed by symmetry, such as a path and its time reverse. 
These will not have cancelling phases and therefore they contribute to $W_{AB}$.
 Let us estimate the correction $\delta \sigma$ to the Drude formula  that the interference term produces. We may think of it as a random quantity and calculate its variance. 
The true conductivity $\sigma = \sigma_{dr} +\delta\sigma$ is proportional to $W_{AB}$ so 
\begin{equation}
\Delta G \propto \frac{1}{N^2}\sum_{ijkl}\mathcal{A}^*_i\mathcal{A}_j\mathcal{A}^*_k\mathcal{A}_l\label{eq:EstimateUCF}
\end{equation}
\begin{figure}
\includegraphics[clip, trim = 80pt 250pt 230pt 50pt, width=\columnwidth]{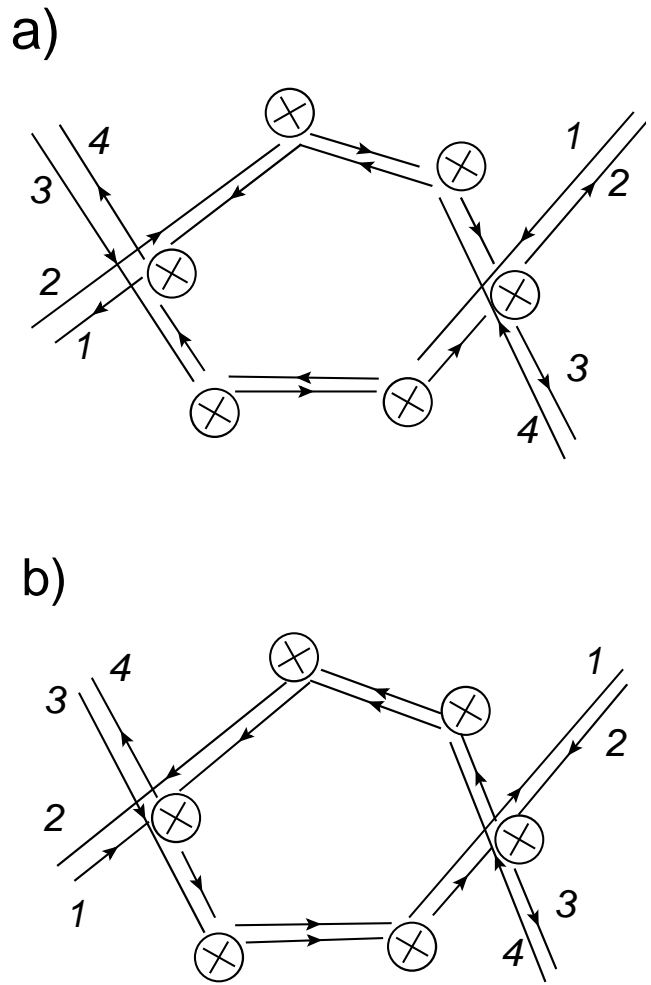}
\caption{\label{fig:PathsUCF}Examples of an interference contribution to the variance of the conductivity. The crossed circles represent static impurities. a) The pair of paths 1 and 2 contribute to the classical probability probability to propagate. Because the two paths are different they have a random phase, which means the sum over all paths is self cancelling. But combined with the paths 3 and 4, the diagram makes a non-vanishing contribution to the variance of the conductivity.  b) A Cooperon contribution, where the path 3 is the time reverse of path 1 and likewise for 4 and 2.
}\end{figure}
There are two sets of paths that give a nonvanishing contribution to \niceref{eq:EstimateUCF}. 
The ``Diffuson"  term where path $i = l$   and $j = k$ and the ``Cooperon" term where path $k$ is the time reverse of path $i$ and likewise for $j$ and $l$. These are illustrated in Fig. \ref{fig:PathsUCF}.
Substituting these paths into \niceref{eq:EstimateUCF}, gives a contribution $\sim \left(\sum_i |\mathcal{M}_i|\right)^2\sim N^2$, not $N$ as in the classical estimate, \niceref{eq:ClassicalG1}. 
This means that the correct expression for $\Delta G$ is independent of the system size.
It follows that this correction is describing processess that occur on linear scales larger that all microscopic lengths and therefore must be universal and independent of material parameters.
 The only possible expression is,
\begin{equation}
\langle \Delta G^2\rangle \sim \left(\frac{e^2}{\hbar}\right)^2. \label{eq:UCFIdeal}
\end{equation}
There are two mechanisms that violate the universality of \niceref{eq:UCFIdeal}: depahsing by inelastic processes characterized by the the inelastic time $\tau_\phi$ (see Refs.~[\onlinecite{Aleiner99,Aleiner02,Altshuler82}] for a detailed discussion of $\tau_\phi$ in mesoscopic fluctuations) and temperature averaging due the dependence of the phases $\mathcal{A}_i$ on the electron energy $\epsilon_i$,
\begin{equation}
\mathcal{A}_i(\epsilon_1)\mathcal{A}_j(\epsilon_2) \propto \exp\left[i\left(\epsilon_1 - \epsilon_2\right)L_i/v_F\right].
\end{equation}
The dephasing restores the central limit theorem in the sense that the system can now be separated into uncorrelated subsystems of size $\ell_\phi \equiv \sqrt{\mathcal{D}\tau_\phi}$. 
Here $\mathcal{D} = v^2_F \tau_{tr}$ is the electron diffusion constant. The temperature averaging similarly means  that contributions from energy differences larger $\epsilon_1-\epsilon_2 \sim \hbar/\tau_\phi$ are independent. This results in
\begin{equation}
\langle\Delta G^2\rangle \sim \left(\frac{e^2}{\hbar}\right)^2\left(\frac{\ell_\phi}{L}\right)^{4-d} \left(\frac{\hbar}{T\tau_\phi}\right),\label{eq:UCFReal}
\end{equation}
where $d$ is the dimensionality of the sample.
 
While $\delta G$ is not directly observable, this correction manifests as the universal conductance fluctuations. 
If an adjustment is made to the system - a change in chemical potential, thermal cycling, magnetic field etc... - the phases in the interference term will be changed and so the interference will be randomized, leading to fluctuations in the conductivity.
 These fluctuations are universal in the sense that they do not depend on physics at the scale $\ell_{imp}$ or $\lambda_F$ , but on much longer scales like the system size or phase coherence length.

Returning to the TLS, we now understand how the motions of the impurities may affect the conductivity. 
\begin{figure}
\includegraphics[width=\columnwidth]{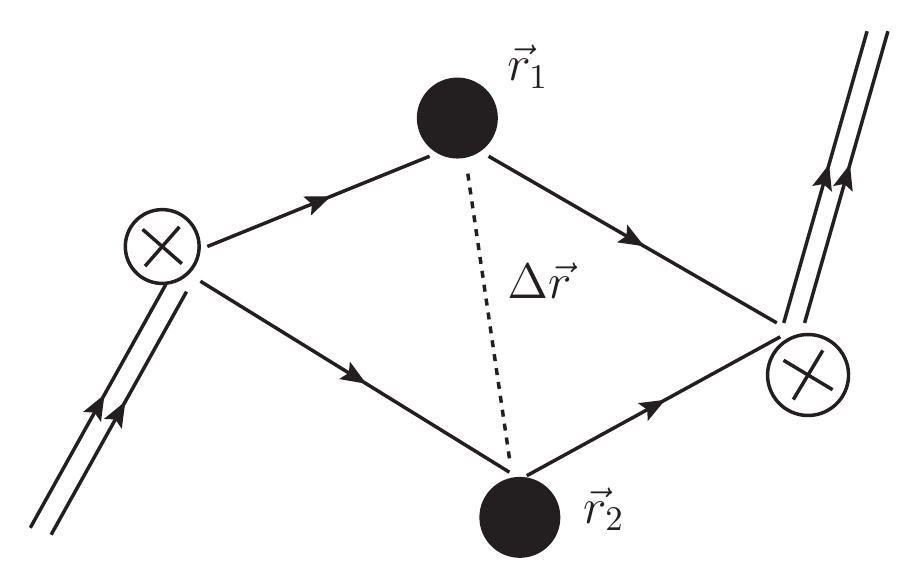}
\caption{
\label{fig:PathsAlpha}
Figure showing the change in the geometric length of a path because of a shift a mobile impurity from position $\vec{r}_1$ to $\vec{r}_2$. The crossed circles represent static impurities and the solid dot shows two possible positions of a TLS.}
\end{figure}
Consider a path involving the scattering on a mobile impurity (TLS) as in Fig. \ref{fig:PathsAlpha}. 
The geometric length of the paths differ depending on the location of the impurity. 
Therefore, the accumulated phase $\phi_i$ of the trajectory depends on the state of the TLS.
We write $\phi_i = p_F L_i +\alpha_{r}$ where $r= {1,2}$ is the state of the TLS. 
The numbers $\alpha_{1,2}$ are effectively random since they depend on the orientation of the electron path and the displacement $\vec{r}_{12}$ between the two sites of the mobile impurity. 
Thus the contribution of the path $i$ to the fluctuation of the conductance becomes dependent on the state of the TLS,
\begin{equation}
\Delta G_{i,r} \sim \cos\left(k_F L_i +\alpha_r\right). \end{equation}
Substituting such paths into \niceref{eq:EstimateUCF} we can calculate the contribution to the conductance fluctuation for paths passing through the TLS. Assuming that $\alpha_1 - \alpha_2 \gg 1$, the sign of $\Delta G_{i,r}$ is random and terms where $r\neq s$ do not contribute. Therefore, [see \niceref{eq:CondPrb}],  
\begin{equation}
\begin{aligned}
\Delta G_{i}(t)\Delta G_{i}(t') &\propto \sum_{r}P^{eq}_{r} P(r,t|r,t')\\
&\propto P^{eq}_1 P^{eq}_2 e^{-t/ \tau_{12}}.
\end{aligned}
\end{equation}

The correlation function of the conductances is determined by the impurity dynamics. The summation over different TLS lead to the correction of \niceref{eq:UCFReal}
\begin{equation}
\begin{aligned}
&\ll\overline{\Delta G(t)\Delta G (t')}\gg \\ \sim &\quad\left(\frac{e^2}{\hbar}\right)^2\left(\frac{\ell_\phi}{L}\right)^{4-d} \left(\frac{\hbar}{T\tau_\phi}\right)\left(\frac{\tau_\phi}{\tau_*}\right)\mathcal{K}(t-t'),\label{eq:NoiseEstimate}
\end{aligned} 
\end{equation}
where $\ll \cdot\gg$ indicates an average over the positions and tunneling rates of the TLS.

The time $\tau_*$ is the elastic scattering time of an electron from a moblie impurity and the factor $\tau_\phi/\tau_* \ll 1$ is the fraction of paths that encounter a mobile impurity before the phase coherence is destroyed. This factor can also be understood as follows. The scattering time $\tau_*$ is approximately the density of states $\nu$ over the density of the TLS, $\rho_*$.  This gives us 
\begin{equation} \left(\frac{\tau_\phi}{\tau_*}\right) = \left(\frac{\rho_*\ell_\phi^d }{g\left(\ell_\phi\right)}\right),\end{equation}
where $g\left(\ell_\phi\right) = \nu\mathcal{D}\ell_\phi^{d-2}$ is the conductance at the scale $\ell_\phi$ in units of $e^2/\hbar$. The phase coherence splits the system into cells of volume $\ell_\phi^d$ each with $\rho_*\ell_\phi^d$ impurities. Therefore to produce a change in the conductance of order $e^2/\hbar$ in a sample of linear size $\ell_\phi$, one must move a number of impurities equal to $g\left(\ell_\phi\right)$. 

We can compare Eqs.~(\ref{eq:DefCalF})~and~(\ref{eq:NoiseEstimate}) by using the facts that on applying a voltage V, the current $I = G(L)V$ and the fluctuations $\delta I = \delta G V$. Further the conductances at scales $L$ and $\ell_\phi$ are related by $G(L) = \frac{e^2}{\hbar}g(\ell_\phi)\left(\frac{\ell_\phi}{L}\right)^{2-d}$. 
We thus obtain a relationship between the functions $\mathcal{F}(t)$ and $\mathcal{K}(t)$,
\begin{equation}
\frac{\mathcal{F}(t)}{\mathcal{K}(t)} \propto \frac{\ell_\phi^d}{g\left(\ell_\phi\right)^2}\left(\frac{\hbar}{T\tau_*}\right).\label{eq:RelTLSNoise}\end{equation}

Equation (\ref{eq:RelTLSNoise}) describes the mechanism of quantum interferance that translates the microscopic motion of the TLS into an observable noise. We will show now that this interference invetiably leads to the memory effect, not previously  studied in the literature. 

\subsection{Memory effect} 
Memory effects are the slow responses of, say, the conductivity $\sigma\left(n_e, B\right)$ to sudden changes of the electron density $n_e$ or the applied magnetic field $B$, as illustrated in Fig. \ref{fig:PlotCondQuench}.
 After the change, the conductivity $\delta\sigma(t)$ is usually larger then its equilibrium value $\sigma_f(n_e+\delta n,B +\delta B)$ and approaches this equilibrium value very slowly, without any visible time scale.
 Moreover, if after some time $t_h$, $n_e$ and $B$ are returned to their starting value, $\sigma$ will jump again (the value and even the sign of the jump depending on $t_h$) and then return to the starting value $\sigma(n_e, B)$ during a time of the order of $t_h$. 

\begin{figure}
\includegraphics[clip, trim = 80pt 400pt 80pt 60pt,width=\columnwidth]{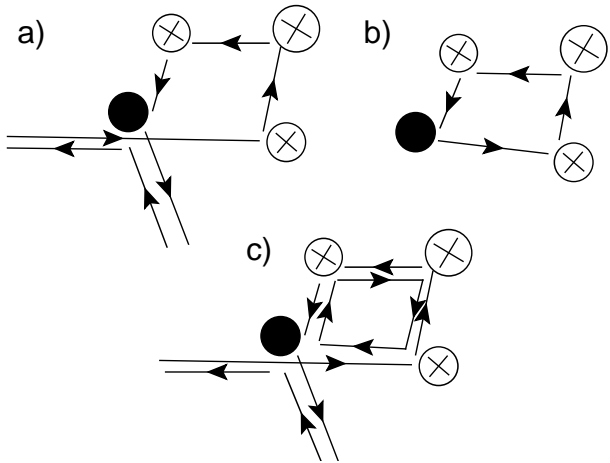}
\caption{\label{fig:PathsMemEffectNew}Semi-classical paths demonstrating the memory effect. The impurity in the TLS is represented by a solid dot and the crossed circles represent static impurities. a) A multiple scattering contribution to the scattering rate of the TLS with a random phase  b) A contribution to the energy in the semi-classical picture. c) An interference contribution to the covariance of the scattering rate and the energy.
}
\end{figure}

We give here a qualitative explanation of this behavior using the concepts introduced in Sec.~\ref{sec:Results}A. The rigorous derivation of these results is relegated to Sec. \ref{sec:MemEffect}.

As before consider the interference contribution to the conductivity from two trajectories shown in Fig.~\ref{fig:PathsMemEffectNew}(a). 
The contribution to the conductivity $\Delta \sigma_i$ from this path corresponds to an enhancement of the scattering rate $1/\tau_{tr}$, and so the effect can be estimated as,

\begin{equation}
\frac{\Delta\sigma_i}{\sigma} \sim -\sum_{r=1,2}\cos(k_F L_i +\alpha_r) P_r \label{eq:delSigma1Path},
\end{equation}
where $P_r$ is the probability for the TLS to be in state $r$. Because the phase of the cosine is random  one might expect \niceref{eq:delSigma1Path} to vanish on averaging. 
However this neglects the possibility that the phase is correlated with $P_r$ and is therefore incorrect. Let us see how this correlation arises.

The equilibrium probability $P^{eq}_r$ for a TLS is given by the Gibbs distribution $P^{eq}_r \propto \exp(-E_r/T)$, where $T$ is the temperature and $E_r$ is the energy of the $r$ state.
 Because the mobile impurity interacts with the electrons, this energy will depend on the density of electrons $\rho(r)$ near the mobile impurity. The density of electrons itself fluctuates throughout the metal because of the Friedel oscillations\cite{Friedel} of the randomly placed impurities. The role of Friedel oscillations in the interaction correction to the conductivity is discussed in Refs.~[\onlinecite{RAG,ZNA}]. Such a fluctuation of the energy $\delta E_r$ will produce a fluctuation in the occupation probability $\delta P_r$,
\begin{equation}
\delta P_1 - \delta P_2  =-\frac{\delta E_1-\delta E_2}{T} P^{eq}_1 P^{eq}_2 .\label{eq:RelateEP}\end{equation}

 Assuming that these density fluctuations are small, we write that the fluctuation of the energy $\delta E_r$ is proportional to the fluctuation of the density $\delta \rho_r$. In the semiclassical picture, the density of electrons at the site $r$ is given by all loops that pass through the site as in Fig~\ref{fig:PathsMemEffectNew}(b), so the path $i$ gives a contribution
\begin{equation}
\delta E^{(i)}_r \propto \delta\rho^{(i)}_r \sim \int\!\! d\epsilon\, n_F\!\left(\epsilon\right) C_{i,r}\left(\epsilon\right), \label{eq:DensFlucSC}\end{equation}
where,
\begin{equation}
 C_{i,r}\left(\epsilon\right) \equiv \cos\left[\left(k_F +\epsilon/v_F\right) L_i + \alpha_r\right],
\label{eq:defCir}
\end{equation}
and $n_F(\epsilon) \equiv \left[1+\exp\left(\epsilon/T\right)\right]^{-1}$ is the Fermi distribution function. 
Substituting Eqs.~(\ref{eq:RelateEP})~and~(\ref{eq:DensFlucSC}) into \niceref{eq:delSigma1Path} and keeping only the non-oscillating terms we obtain,
\begin{equation}
\frac{\Delta \sigma_i} {\sigma_D} \sim -\frac{1}{T}\int\! d\epsilon\, n_F\!\left(\epsilon\right) \cos\left(\frac{\epsilon}{v_F}L_i\right)P^{eq}_1 P^{eq}_2.\label{eq:delSigma1Path2} \end{equation}

The next step is the summation of \niceref{eq:delSigma1Path2} over all the diffusive paths that involve the scattering off of the mobile impurities. 
This is precisely the sum [\niceref{eq:EstimateUCF}] we have discussed in Sec. 2A, where we found that the change in the conductance is given by the inverse conductance on the scale $\ell_\phi$.
 The only difference is that, because of the integral over $\epsilon$ in \niceref{eq:delSigma1Path2}, the phase coherence will already be destroyed for paths longer than $\hbar v_F/T$. 
This corresponds to a diffusive length $L_T = \sqrt{\hbar\mathcal{D}/T}\ll \ell_\phi$ [see Ref.~[\onlinecite{AAK}]]. Calling the total correction to the conductivity $\sigma_*$, we obtain that
\begin{equation}
\frac{\sigma_*}{\sigma_D} \approx -\frac{1}{g\left(L_T\right)}\frac{1}{T\tau^*}.\label{eq:defSigmaStar}\end{equation}
Equation~(\ref{eq:defSigmaStar}) is a quantum correction to the conductivity with a singular dependence on temperature. Similar effects were discussed in Ref.~[\onlinecite{KozubRudin97}] in relation to zero bias anomalies in point contacts. 

Due to the small factor $1/\left(T \tau_*\right)$ this correction is not observable in bulk systems in comparison with the interaction correction\cite{AltshulerAronov85}. It is only the memory effect that makes the correction \niceref{eq:defSigmaStar} observable.

Let us  at time $t=0$ suddenly change the electron density so that $k_F\rightarrow k_F'$, or apply a magnetic field $B$. The electrons equilibriate instantly compared to the time scales we are interested in, so we should change in  \niceref{eq:delSigma1Path} and \niceref{eq:defCir}
\begin{equation}
C_{i,r}\left(\epsilon\right) \rightarrow \tilde{C}_{i,r}(\epsilon)\equiv\cos\!\left(2\pi\frac{\Phi_i}{\Phi_0}\right)\!\cos\left[\!\left(\! k_F' +\!\frac{\epsilon}{v_F}\!\right)\!L_i+\alpha_r\!\right]\!,
\end{equation}
where $\Phi_i$ is the flux enclosed by the diffusive path and $\Phi_0 = hc/e$ is the flux quantum. However, the occupation probability of a TLS does not immediately follow the change in density, because it relaxes only on the long time scale $\tau_{12}$. Therefore, we should write for the occupation probability,
\begin{equation}
\begin{aligned}
\Delta P_r(t) = &-\frac{e^{-t/\tau_{12}}}{T}\int d\epsilon n_F\left(\epsilon\right)C_{i,r}\left(\epsilon\right)\\
&-\frac{1- e^{-t/\tau_{12}}}{T}\!\int\!\! d\epsilon\, n_F\!\left(\epsilon\right)\tilde{C}_{i,r}\left(\epsilon\right).\\
\end{aligned}
\end{equation}  
Then, \niceref{eq:delSigma1Path} yields 
\begin{equation}
\begin{aligned}
\frac{\Delta \sigma_i\!\!\left(t\right)}{\sigma} &\sim -\!\!\sum_{r = 1,2}\!\!\!P_1 P_2\!\int\!\!d\epsilon\, n_F\left(\epsilon\right)\! \Big[\frac{e^{-t/\tau_{12}}}{T}C_{i,r}\left(\epsilon\right)\tilde{C}_{i,r}\left(0\right)\\ &- \frac{1- e^{-t/\tau_{12}}}{T}\tilde{C}_{i,r}\left(\epsilon\right)\tilde{C}_{i,r}\left(0\right)\Big].
\end{aligned}
\end{equation}
Once again, keeping only the terms which do not oscillate on the scale of $1/k_F$ we obtain instead of \niceref{eq:defSigmaStar}
\begin{equation}
\begin{aligned}
\frac{\Delta \sigma_i\!\!\left(t\right)}{\sigma} &=-\frac{P_1 P_2}{T}\\ \times\int\!\!d\epsilon\,\bigg\{&e^{-t/\tau_{12}} \!\cos\left[\left(\!k'_F-k_F+\frac{\epsilon}{v_F}\!\right)\!L_i\!\right]\cos \frac{2\pi \Phi_i}{\Phi_0}\\
& +\left(1-e^{-t/\tau_{12}}\right)\cos \frac{\epsilon}{v_F} \cos^2 \frac{2\pi \Phi_i}{\Phi_0}\bigg\}.
\label{eq:delSigma1Path3}
\end{aligned}
\end{equation}
\fullref{eq:delSigma1Path3} is the key for the qualitative understanding of the memory effect. 
The first term characterizes the slow decay of the system's memory of the initial interference pattern. 
The second term characterizes the slow approach of the conductivity to the new equilibrium. 
The term $\text{cos}^2\left( 2\pi \Phi_i/ \Phi_0\right)$ describes the suppression of the constructive interference between time-reversed paths by the magnetic field. The same suppression by magnetic field appears in the $1/f$ noise\cite{Birge90,TrionfiNatelson} and is evidence of the importance of mesoscopic physics in the system.

\begin{figure}
\includegraphics[width=\columnwidth]{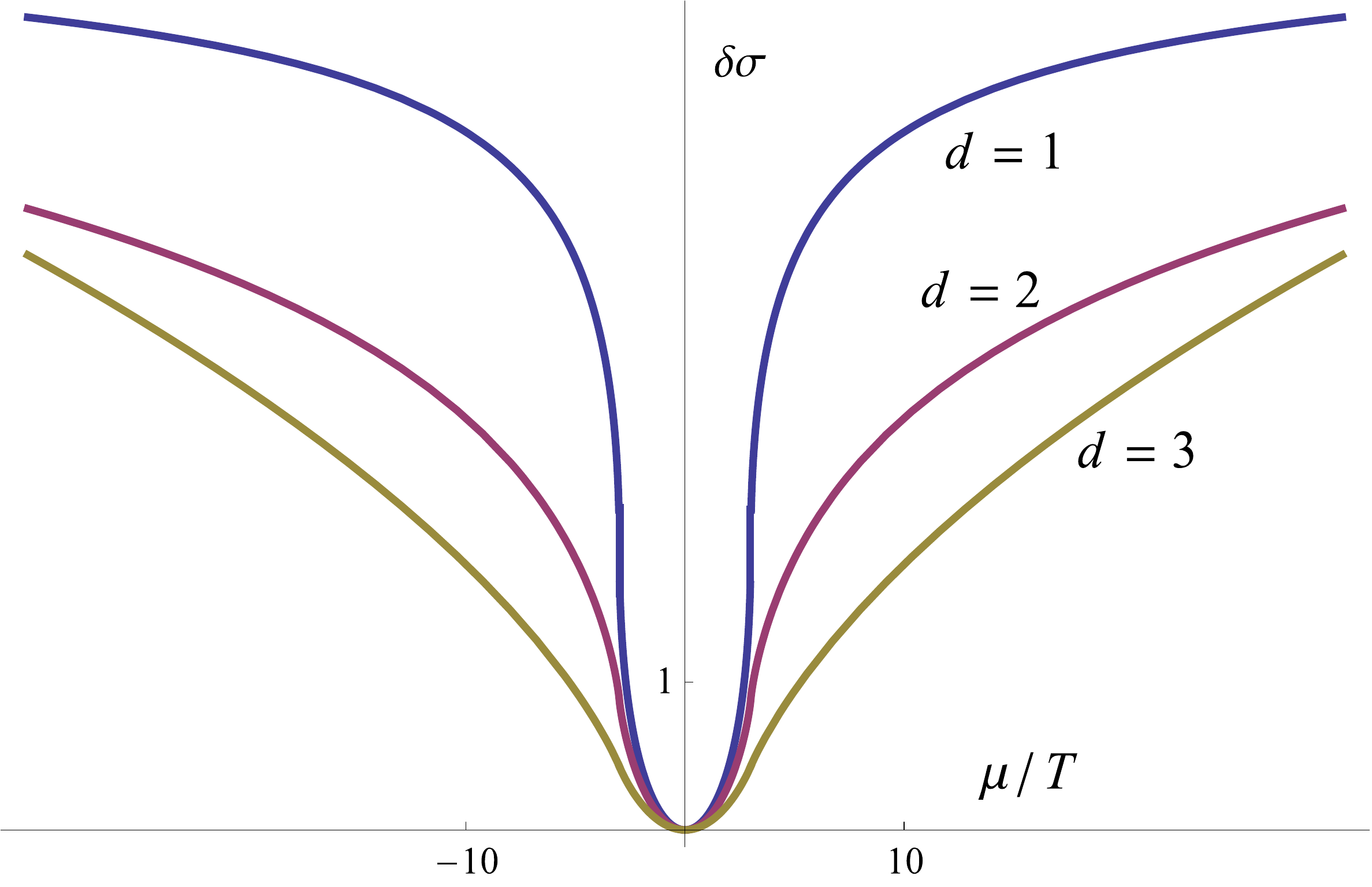}
\caption{\label{fig:PlotZBA}Graph of the zero-bias anomaly in the conductivity. The conductivity and chemical potential are measured from the resting values. The curves are obtained by numerical integration of \niceref{eq:LineShape}
}
\end{figure}

\fullref{eq:delSigma1Path3} has several immediate applications. 
Let us  consider the change in conductivity immediately after a change in the density. \footnote{Note the interaction correction does not produce any singular density dependence, because the self consistent potential created by the electron-electron interactions equilibriates almost instantaneously.} Summing over all the trajectories and all the TLSs in \niceref{eq:delSigma1Path3} we obtain the total correction to the conductivity,
\begin{equation}
\frac{\delta \sigma\!\left(B,k_F',t =0\right)}{\sigma_D} = -\frac{1}{g\left(L_T\right)}\frac{1}{T\tau_*} S\left(\frac{v_F|k_F-k_F'|}{T},\frac{L_T}{L_B}\right),
\label{eq:ZBA}
\end{equation}
where $L_B \equiv \sqrt{\hbar c/\left(eB\right)}$ is the magnetic length and the function $S(x,y)$ counts the fraction of diffusive
paths whose interference is not destroyed due to changes in $k_F$ or $B$. It has the asymptotic limits
\begin{equation}
S\left(0,0\right) = 1;\,S\left(x\rightarrow\infty,y\right) = S\left(x,y\rightarrow \infty\right) = 0.
\end{equation}
The explicit form of $S$ is given in \niceref{eq:MemEffectMain}. The dependence of the conductivity on the density is shown in Fig.~\ref{fig:PlotZBA}. It can be seen as a fingerprint of the electron density that is stored in the TLS.

\begin{figure}
\includegraphics[width=\columnwidth]{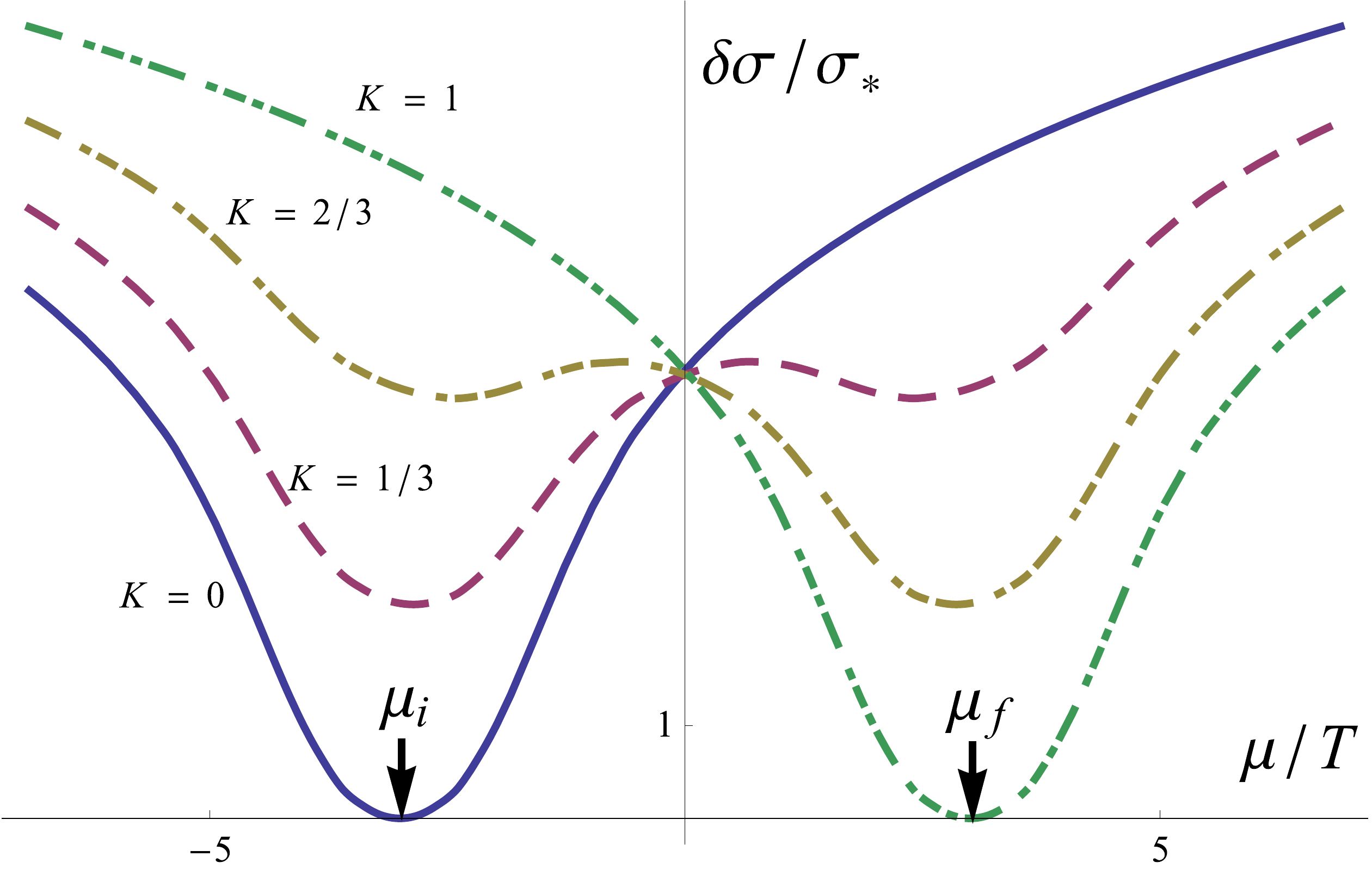}
\caption{\label{fig:PlotCondRelax}Graph showing the relaxation in a thin film of the conductivity singularity from the old Fermi level $\mu_i$ to the new Fermi level $\mu_f$. The curves are labelled by the fraction $K$ of TLS that have relaxed to the new equilibrium. 
}
\end{figure}

The time dependence of the conductivity is even more dramatic. Taking \niceref{eq:delSigma1Path3} and summing over all the diffusive paths and all the TLSs with the distribution function from \niceref{eq:DefCalK} we obtain,
\begin{equation}
\begin{aligned}
&\frac{\delta\sigma\!\left(B,k_F'; t\right)}{\sigma_D}\!=\! -\frac{1}{g\left(L_T\right)}\frac{\hbar}{T\tau_*}\bigg\{\!\mathcal{K}(t) S\!\left(\!\frac{v_F|k_F-k_F'|}{T};\frac{L_T}{L_B}\right)\\&\quad +\frac{1}{2}\left(\mathcal{K}(0) -\mathcal{K}(t)\right)\left[1+S\left(0,\sqrt{2}\frac{L_T}{L_B}\right)\right]\bigg\}.
\end{aligned}
\end{equation}
This dependence has two anomalies, one at the old Fermi level and the second at the new Fermi level. The ratio between the amplitude of these anomalies characterizes the fraction of the TLS that have adjusted to the new electron density. The form of the density dependence is shown on Fig. \ref{fig:PlotCondRelax}.

The function $\mathcal{K}$ is precisely the function given in \niceref{eq:DefCalK} which determines the correlations of the $1/f$ noise [see Eqs.~(\ref{eq:DefCalF})~and~(\ref{eq:RelTLSNoise})].
 Moreover, the unknown factor $\hbar/\left(T\tau_*\right)$ is removed if the memory effect is expressed in terms of the measurable correlation function of the $1/f$ noise from \niceref{eq:DefCalK},
\begin{equation}
\begin{aligned}
&\frac{\delta\sigma\left(B,k_F'; t\right)}{\sigma_D}= -\frac{1}{V_q}\bigg\{\mathcal{F}(t) S\left(\frac{v_F|k_F-k_F'|}{T};\frac{L_T}{L_B}\right)\\&\quad +\frac{1}{2}\left(\mathcal{F}(0) -\mathcal{F}(t)\right)\left[1+S\left(0,\sqrt{2}\frac{L_T}{L_B}\right)\right]\bigg\},
\end{aligned}
\end{equation}
where $V_q$ is the effective volume of the subsystem which contribute to the memory effect and is defined by,
\begin{equation}
\frac{1}{V_q} \equiv \frac{g\left(\ell_\phi\right)^2}{\ell_\phi^d g\left(L_T\right)} \approx \nu T \left(\tau_\phi T\right)^{2-d} \label{eq:DefVq}.\end{equation}
The time $\tau_\phi$ can be extracted from the usual weak localization magneto-resistance measurement.

\begin{figure}
\includegraphics[width=\columnwidth]{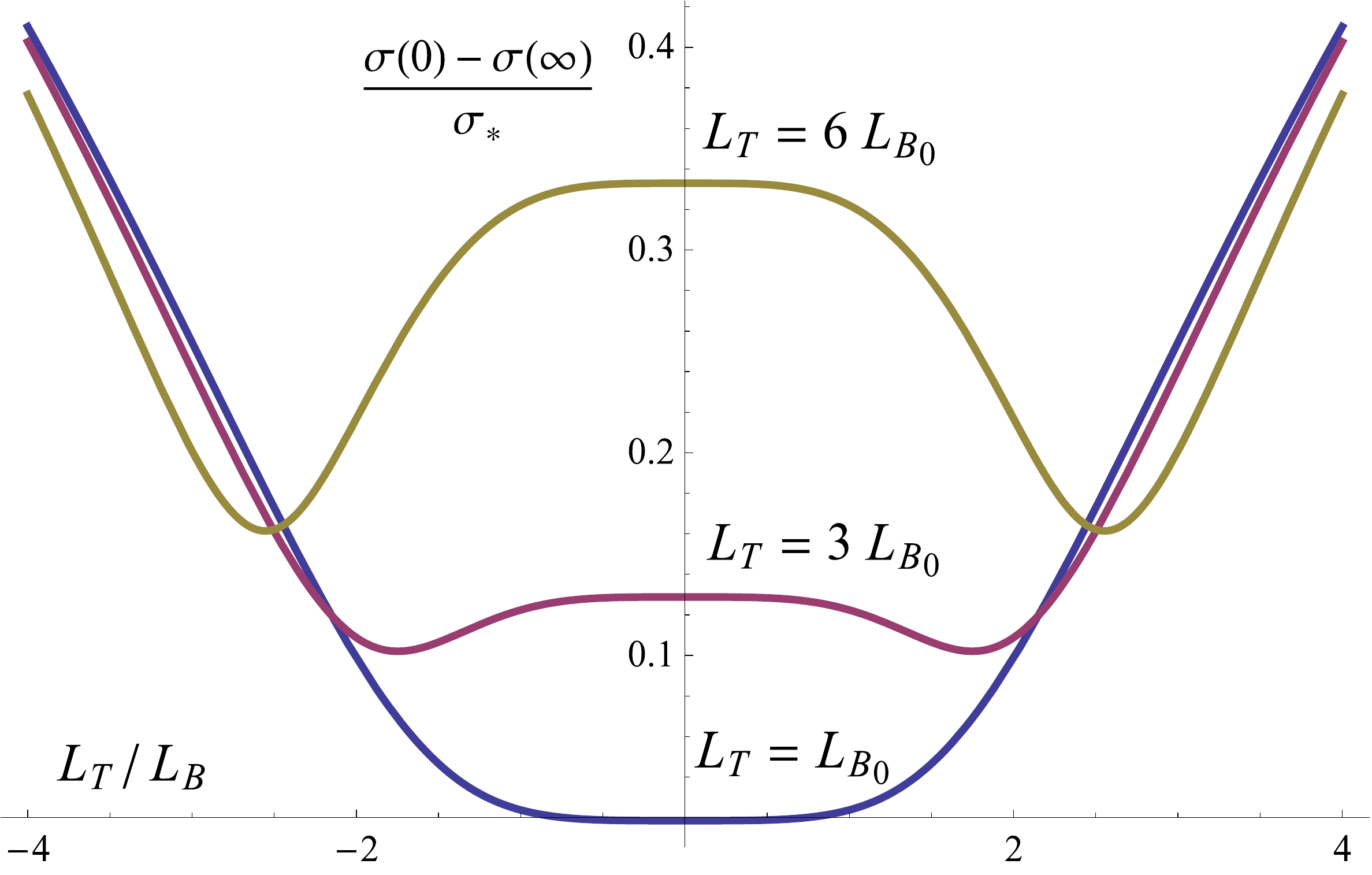}
\caption{\label{fig:PlotMagMem}Plot of the magnetic memory effect. The curves plot the difference between $\sigma(0)$, the conductivity of a sample equilibriated in zero-field, and $\sigma(\infty)$, the conductivty after the sample has equilibriated in a transverse field with magnetic length $L_{B_0}$. The curves are shown for different choices of the resting magnetic length $L_{B_0}$ and plotted in term of the $L_B$, the magnetic field length when the conductivity is measured.  They are obtained by numerical evaluation of \niceref{eq:MemEffectScaling}.}
\end{figure}

The closest relative of the density memory effect discussed above is the magnetic field memory effect. 
Let us keep the density fixed and switch the magnetic field at $t=0$ from $B=0$ to $B=B_0$.  
Then at some later time $t$ we briefly shift the magnetic field to a third value $B$ and measure the resistance. Repeating the arguments starting from  \niceref{eq:delSigma1Path3} we find the that the time-dependent part\footnote{There is also a contribution from the anamolous magneto-resistance, but this does not depend on time} of the resistance is,
\begin{equation}
\begin{aligned}
&\frac{\delta\sigma\left(B,k_F'; t\right)}{\sigma_D}= -\frac{1}{V_q}\bigg\{\mathcal{F}(t) S\left(0;\frac{L_T}{L_B}\right)\\&\quad +\frac{1}{2}\left(\mathcal{F}(0) -\mathcal{F}(t)\right)\left[S\left(0,\frac{L_T}{L_{B_+}}\right)+S\left(0,\frac{L_T}{L_{B_-}}\right)\right]\bigg\},\\
&L_{B_\pm} \equiv \sqrt{\frac{\hbar c}{e\left|B_0\pm B\right|}}.
\end{aligned}
\label{eq:PlotMagMem}
\end{equation}
At large value of the magnetic field ($2L_T \gtrsim  L_B$) the magneto-resistance shows a distinct two dip structure, shown in Fig. \ref{fig:PlotMagMem}. Note that the magneto-resistance is always symmetric. 
This is because the electrons are always in quasi-equilibrium and so Onsager's relation applies.

There is a different way to probe the same memory physics, by performing a cyclic perturbation of the system. 
We can at $t=0$ turn on a magnetic field or change the density and wait for a time $t_h$. We then switch off the magnetic field or return the density to its previous value. We may then measure the conductivity $\sigma(t)$ at time $t>t_h$, when the system has the parameters as at $t<0$ but still retains a memory of the period $0<t<t_h$. 
This protocol corresponds to the correction of the energy levels of the TLS only during the finite time $t_h$. We obtain instead of \niceref{eq:delSigma1Path} at $t>t_h$,
\begin{equation}
\begin{aligned}
\delta P_r(t) &= \int\!d\epsilon\, n_F\left(\epsilon\right)\left[\tilde{C}_{i,r}\left(\epsilon\right) - C_{i,r}\left(\epsilon\right)\right]\\ &\times\left(e^{-t/\tau_{12}} - e^{-\left(t-t_h\right)/\tau_{12}}\right).
\end{aligned}
\end{equation}
Repeating the previous derivation we obtain a correction to the conductivity,
\begin{equation}
\frac{\delta \sigma(t)}{\sigma_D} = \frac{\mathcal{F}\left(t\right) -\mathcal{F}\left(t-t_h\right)}{V_q}\!\left[1\!-\!S\!\left(\frac{v_F|k_F -k_F'|}{T},\frac{L_T}{L_B}\right)\right]\!.
\label{eq:DoubleSwitch}
\end{equation}
Equation~(\ref{eq:DoubleSwitch}) describes the relaxation dynamics of the conductivity. 
This protocol has the advantage of being insensitive to the fastest time of scale of the TLS dynamics [it does not contain $\mathcal{F}(0)$].
 It is also non-invasive in that it does not require sweeps of the parameters which may affect the evolution of the system. However the measurement of $\Delta \sigma(t)$ and the jumps in conductivity can still be used to extract the function $S$. Therefore the consistency of the different protocols would be an important test of this framework.

We conclude this section by noting that the theory developed here can predict the change in conductivity from any history of the density or magnetic field, by application of \niceref{eq:MemEffectMain}. 
It therefore constitutes a complete description of the memory phenomenon.

\section{Diagrammatics for electrons and TLS\label{sec:Diagrammatics}}
In this section we will introduce the  diagrammatic technique for disordered metals with TLS and perform a rigorous derivation of the results discussed in Sec.~\ref{sec:Results}.
 The model is defined in Subsections~\ref{sec:model}~and~\ref{sec:FlucDiss}. 
Subsections~\ref{sec:UCF} and~\ref{sec:1f} rederive the known results for the mesoscopic fluctuations and the $1/f$ noise in order to harmonize the notation and allow an easy comparison with the memory effect. 
The quantitative derivation of the memory effect is performed in subsection~\ref{sec:MemEffect}.

We make several simplifying assumptions, but they do not appear crucial to the results: 
(i) all dependence on the electron-electron and electron-phonon interactions appears only through the phase coherence length $\ell_\phi$,
 (ii) we work to leading  in $g\left(\ell_\phi\right)^{-1}$, (iii) we work to leading order in $T\tau_\phi/\hbar\ll 1$, (iv) the calculation is perturbative in the density of the TLS. We set $\hbar=c = 1$ in all intermediate formulae.

\subsection{Model\label{sec:model}}
The total Hamiltonian for our system is
\begin{equation}
\hat{H}= \hat{H}_{metal} + \hat{H}_{TLS} +\hat{H}_{el-TLS}.\label{eq:totHam}
\end{equation}
 The metallic system is described by the Hamiltonian, 
\begin{equation}
\hat{H}_{metal} = \int \!d^d \vec{r}\,\,\psi^\dagger\left(\vec{r}\right)\!\left[\epsilon\left(-i\vec{\nabla} -e\vec{A}\right) + U\left(\vec{r}\right)\right]\!\psi\left(\vec{r}\right).
\end{equation}

Here $\psi^\dagger$ is the electron creation operator, $\epsilon(p)$ is the electron spectrum, $\vec{A}$ is the vector gauge potential, $U(r)$ is a random scalar field representing static disorder and we suppress throughout spin indices. We take the simplest model of a local Gaussian disorder with correlation function
\begin{equation}
\ll U(\vec{r}) U(\vec{r}')\gg = \frac{1}{2\pi\nu\tau}\delta^{(d)}(\vec{r}-\vec{r}').\label{eq:ImpurityGaussian}
\end{equation}
Here $\nu$ is the electron density of states per spin at the Fermi level and $\tau$ is the scattering rate. The double brackets $\ll \cdot\gg$ throughout this text mean average over both the static impurities and all others kinds of disorder.

The Hamiltonian for the TLS,
\begin{equation}
\hat{H}_{TLS} = \sum^{N_{TLS}}_{i=1} \hat{h}_i,\end{equation}
is a sum of Hamiltonians for each of the $N_{TLS}\gg 1$ two level systems,
\begin{equation}
\hat{h}_i = \Delta_m\left[ x_i\hat{\sigma}_z^i + e^{-r_i}\hat{\sigma}_x^i\right].\label{eq:defhi}
\end{equation}
The $\hat{\sigma}_{x,y,z}^i$ are the usual Pauli matrices, commuting for different TLS. The parameters $x_i$ are indepedent random variables uniformly distributed $0\leq x_i\leq1$, and  $r_i$ are indepedent random variables uniformly distributed $0 \leq r_i \leq R$, where the large distance cutoff $R\gg1$ characterizes the lowest frequency at which the $1/f$ noise is observed. The energy $\Delta_m$ is the maximal level splitting of a TLS.

The motion of the TLS changes the potential for electrons in the system. As the static potential is already disordered, the efect of the TLS can be modeled as a change of the correlation function of the disordered potential (\ref{eq:ImpurityGaussian}),
\begin{equation}
\hat{H}_{el-TLS} = \int \!\!d^d\vec{r}\,\,V\!\left(\vec{r}; \left\{\hat{\sigma}_i\right\}_{i=1}^{N_{TLS}}\right)\psi^\dagger\!\left(\vec{r}\right)\psi\!\left(\vec{r}\right),
\end{equation}
\begin{equation}
\begin{aligned}
\ll V\left(\vec{r}; \left\{\hat{\sigma}_i\right\}_{i=1}^{N_{TLS}}\right)&\otimes V\left(\vec{r}';\left\{\hat{\sigma}_i\right\}_{i=1}^{N_{TLS}}\right)\gg\\& = \frac{\gamma}{2\pi\nu\tau}\delta^{(d)}\left(\vec{r}-\vec{r}'\right)\sum_{i=1}^{N_{TLS}} \hat{\sigma}_z^i\otimes\hat{\sigma}_z^i,\end{aligned}\end{equation}
where $\gamma \ll 1$ describes the ratio of scattering of the mobile impurities to the elastic scattering. It is important to emphasize that averaging here is performed only over the spatial locations of the TLS and that the average  over the parameters of the TLS ($x_i$ and $r_i$) should be performed in the final answer. The resulting diagrammatics are summarised in Fig~\ref{fig:ElecDefs}. 
\subsection{Fluctuation-dissipation theorem for dilute TLS\label{sec:FlucDiss}}

By using the fluctuation-dissipation theorem we may relate the noise and the quantum memory effects without any appeal to the microscopic details of the TLS. For dilute TLS (meaning that the average number of TLS per coherent volume $\ell_\phi^d$ is much less than one) the dynamics of the different TLS are independent. The fluctuations are expressed in the exact Keldysh Green's function,
\begin{equation}
F^K\!\!\left(t_1 -t_2\right) = \frac{1}{2N_{TLS}}\!\!\sum_{i=0}^{N_{TLS}}\!\left\langle \hat{\sigma}_z^i(t_1)\hat{\sigma}_z^i(t_2) + \hat{\sigma}_z^i(t_2)\hat{\sigma}_z^i(t_1)\right\rangle,\label{eq:DefFK}
\end{equation}
Here $\hat{\sigma}^i_z(t)$ is the operator defined in \niceref{eq:defhi} in the Heisenberg representation and the quantum mechanical expectation $\langle\cdot\rangle$ is performed over the equilibrium density matrix of the electron system.  The response of the TLS to the change in it's enviroment, such as perturbations of the electrons, is encoded in the retarded Green's function,
\begin{equation}
F^R\!\!\left(t_1 -t_2\right) = \frac{i}{2N_{TLS}}\sum_{i=0}^{N_{TLS}}\left\langle\left[\hat{\sigma}_z^i(t_1),\hat{\sigma}_z^i(t_2)\right]\right\rangle\theta\left(t_1-t_2\right),
\end{equation}
where $\theta(t)$ is the step function. Note that we remove a factor of $i$ from \niceref{eq:DefFK} so that both $F^K$ and $F^R$ are real functions. 

Further microscopic calculation is relegated to Appendix \ref{sec:TLS}. For our purposes it is sufficient to use the fluctuation dissipation theorem. From the fact that all time scales are much longer than $\hbar/T$ we may write,
\begin{equation}
F^R(t) = \frac{\theta(t)}{T}\frac{\partial F^K(t)}{\partial t}\label{eq:FlucDiss}.\end{equation}

Therefore everything may be expressed in terms of $F^K(t)$. 

\subsection{Mesoscopic conductance fluctuations\label{sec:UCF}}
\begin{figure}
\Large
\makebox[\columnwidth][l]{\raisebox{30pt}{a)}\hspace{-2pt}\includegraphics[width =75pt]{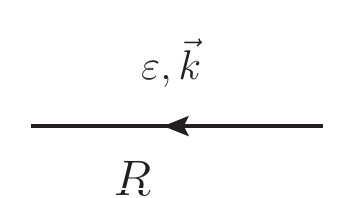} \raisebox{10pt}{$= \left(\epsilon -\varepsilon(\vec{p}) +i0^+\right)^{-1}$}}\newline
\makebox[\columnwidth][l]{b)\raisebox{-3pt}{\includegraphics[width = 75pt]{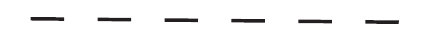}$=\frac{1}{2\pi\nu\tau}$}\hspace{\fill}}\newline
\makebox[\columnwidth][l]{c)\raisebox{-10pt}{\includegraphics[width = 200pt]{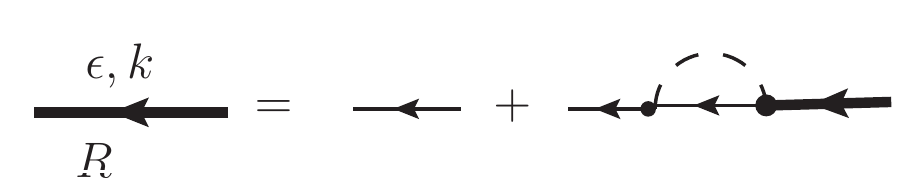}}} $=G^R\left(\epsilon,\vec{p}\right)
=\left(\epsilon - \varepsilon(\vec{p})+\frac{i}{2\tau}\right)^{-1}$\newline
\vspace{5pt}
\makebox[\columnwidth][l]{\raisebox{30pt}{d)}\includegraphics[width = 170pt]{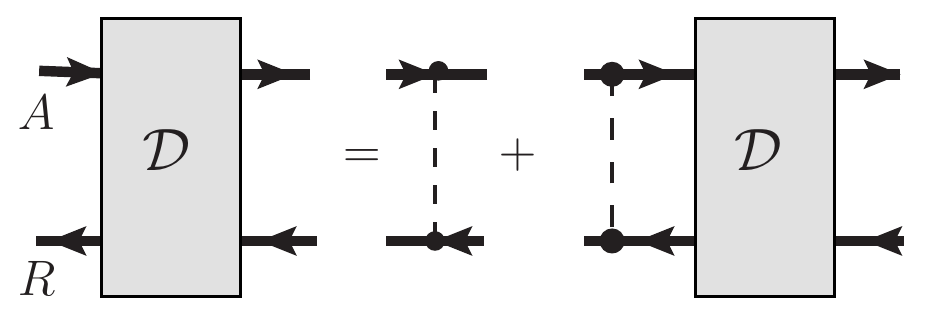}\raisebox{25pt}{$=\frac{1}{2\pi\nu\tau^2}\mathbf{D}$}}
\makebox[\columnwidth][l]{\raisebox{30pt}{e)}\includegraphics[width = 170pt]{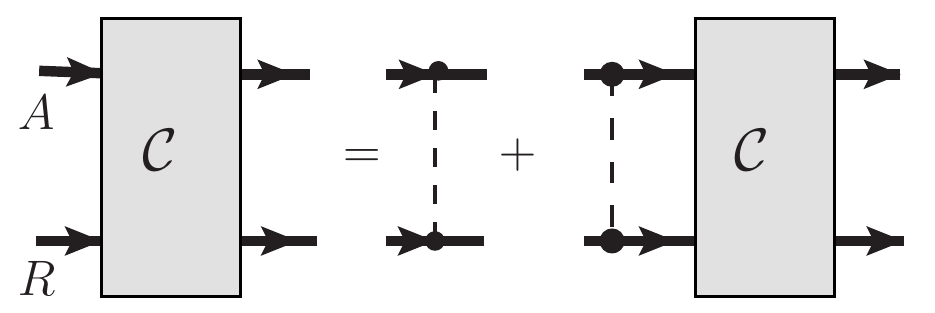}\raisebox{25pt}{$=\frac{1}{2\pi\nu\tau^2}\mathbf{C}$}}
\caption{
\label{fig:ElecDefs}The definition of the diagrammatic elements (a) bare electron Green's function (b) static impurity (c) dressed electron Green's function (d) and (e) the resummation for the Cooperon and Diffuson pole. The external fermion lines are amputated and the functions $\mathbf{D}$ and $\mathbf{C}$ are defined in \niceref{eq:Diffuson} and \niceref{eq:Cooperon}
}\end{figure}
The properties of the conductance fluctuations are well studied. We reproduce the results in this section in order to establish the notation and the building blocks of the diagrammatic technique.  The diagrams for the impurity averaged Green's functions $\ll G^{R,A}\gg$ and the average of their product $\ll G^R G^A\gg$ are shown in Fig. \ref{fig:ElecDefs}. Because we are averaging measurements made at well-separated times we can attach a definite time to each electron line. The most interesting part of the long range dynamics is encoded in the Diffuson and Cooperon propogators $\mathbf{D}$ and $\mathbf{C}$, see Fig.~\ref{fig:ElecDefs}(d,e). These are the solutions of the ``classical" equations, 
\begin{subequations}
\begin{equation}
\begin{aligned}
\bigg\{i\eta  + \Big[i\nabla_{r_1} +&\big(\vec{A}(t_1,r_1)+ \vec{A}(t_2,r_1)\big)\Big]^2 +  \tau_\phi^{-1}\bigg\}\\ \times& \,\, \mathbf{C}\left(\eta,r_1,r_2;t_1,t_2\right) = \delta^{(d)}\left(r_1-r_2\right),\label{eq:Cooperon}
\end{aligned}
\end{equation}
and
\begin{equation}
\begin{aligned}
\bigg\{i\eta + \Big[i\nabla_{r_1} +&\big(\vec{A}(t_1,r_1)- \vec{A}(t_2,r_1)\big)\Big]^2 +  \tau_\phi^{-1}\bigg\}\\ \times & \,\,\mathbf{D}\left(\eta,r_1,r_2;t_1,t_2\right) = \delta^{(d)}\left(r_1-r_2\right),\label{eq:Diffuson}
\end{aligned}
\end{equation}
\end{subequations}
where $\eta\equiv \epsilon_1-\epsilon_2$ is the difference of the energy of the two electron lines. The constant $\tau_\phi$ is the phase coherence time, which captures the effect of the interacting processes not explicitly included in our model, such as phonons. The gauge is fixed with $A^0 = 0$ so that $\mathbf{C}\left(r,r;t_1,t_2\right)$ is invariant under the residual, time-independent gauge transformations.

In the absence of a magnetic field, there is no dependence on the times $t_1$ and $t_2$ and the Fourier transform of the propogators is given by,
\begin{equation}
\mathbf{C}\!\left(\eta,\vec{Q}\right)= \mathbf{D}\!\left(\eta,\vec{Q}\right) = \left(-i\eta + \mathcal{D}Q^2 + \tau_\phi^{-1}\right)^{-1}\label{eq:CooperonB0}.\end{equation}
\begin{figure}
\large
\makebox[\columnwidth][l]{\raisebox{30pt}{a)}\includegraphics[width=75pt]{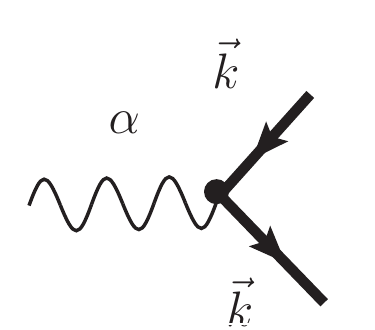}\raisebox{20pt}{$\displaystyle=e\frac{\partial \varepsilon}{\partial p_\alpha}\bigg|_k\equiv ev^\alpha(k)$}}
\makebox[\columnwidth][l]{b)\includegraphics[width=220pt]{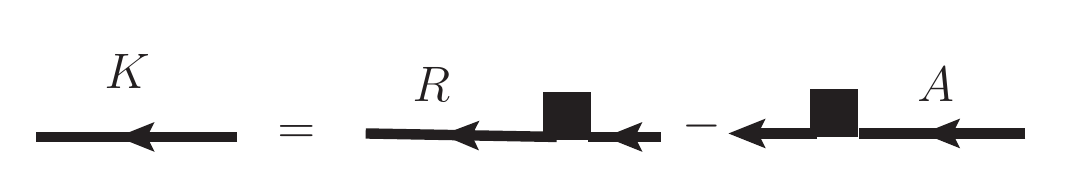}}\newline
\vspace{4pt}
\makebox[\columnwidth][l]{c)\raisebox{-5pt}{\includegraphics[width=30pt]{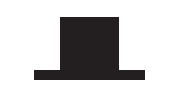}} $=1-2n\left(\epsilon,\vec{p}\right);$}\newline
$\,\,n\left(\epsilon,p\right) =\left(1-\Gamma_\alpha v^\alpha(p)\frac{\partial}{\partial \epsilon}\right)f_F(\epsilon)$\newline
\vspace{5pt}
\makebox[\columnwidth][l]{\raisebox{20pt}{d) $\displaystyle\ll\!\! j_\alpha\!\!\gg\, =(-i)\!\!$ }\includegraphics[width=85pt]{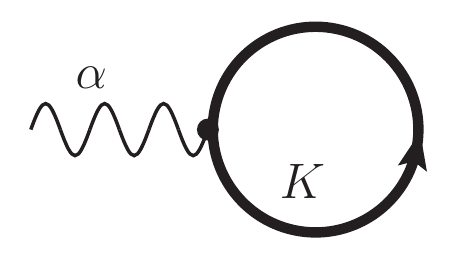}}
\makebox[\columnwidth][l]{\hspace{35pt}$\displaystyle \quad\,= \!2e\!\!\int\!\!d\epsilon\! \left(-\frac{\partial f_F}{\partial \epsilon}\right)\nu\left\langle v^\alpha v^\beta\right\rangle_{FS}\Gamma_\alpha$}
\caption{
\label{fig:DistDefs}The definition of the diagrammatic elements (a) current operator, (b) Keldysh Green's function, (c) electron distribution function, (d) expectation of the current operator. The factor of two comes from the spin summation. The Fermi function $f_F(\epsilon) \equiv \left[1+\exp\left(\epsilon/T\right)\right]^{-1}$. Note the factor of $-i$ in the definition (c) of the average current}
\end{figure}

The non-equilibrium distribution of the electronic system due to a finite current is expressed by the Keldysh Green's function $G^K$ shown in Fig. \ref{fig:DistDefs}(b) or equivalently by the electron distribution function $n\left(\epsilon, \vec{p}\right)$.  The average current, shown in Fig. \ref{fig:DistDefs}(d) reproduces the usual Drude formula. 

\begin{figure}
\large
\vspace{4pt}
\makebox[\columnwidth][l]{\raisebox{50pt}{a)}\includegraphics[width=180pt]{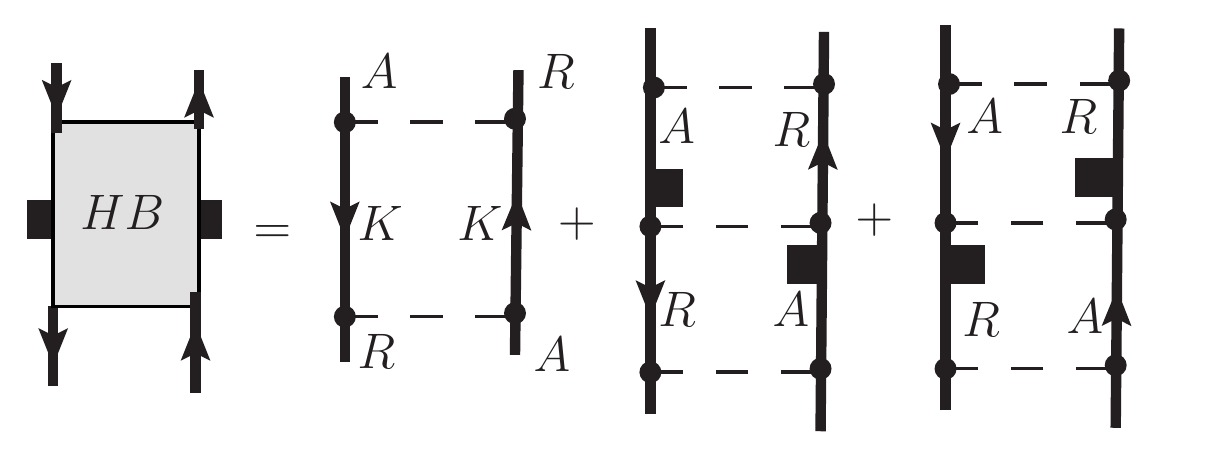}}\newline
\makebox[\columnwidth][l]{\hspace{10pt}$\displaystyle =-\frac{4\pi\tau^2}{ \nu \mathcal{D}}\ll\!\!\vec{j}\!\!\gg\cdot\ll\!\! \vec{j}\!\!\gg f'_F(\epsilon_1)f'_F(\epsilon_2)$}
\makebox[\columnwidth][l]{\raisebox{50pt}{b)}\includegraphics[width=100pt]{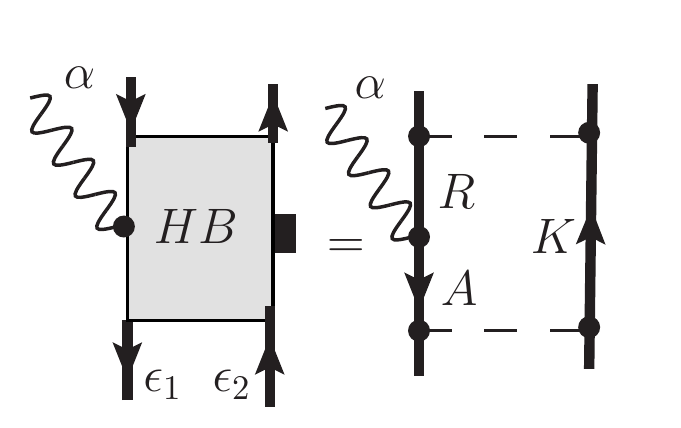}\raisebox{25pt}{$\displaystyle  =4\pi i\tau^2\!\!\ll\!\! j_\alpha\!\!\gg\!\! f'_F(\epsilon_2)$}}\newline
\makebox[\columnwidth][l]{\raisebox{50pt}{c)}\includegraphics[width=110pt]{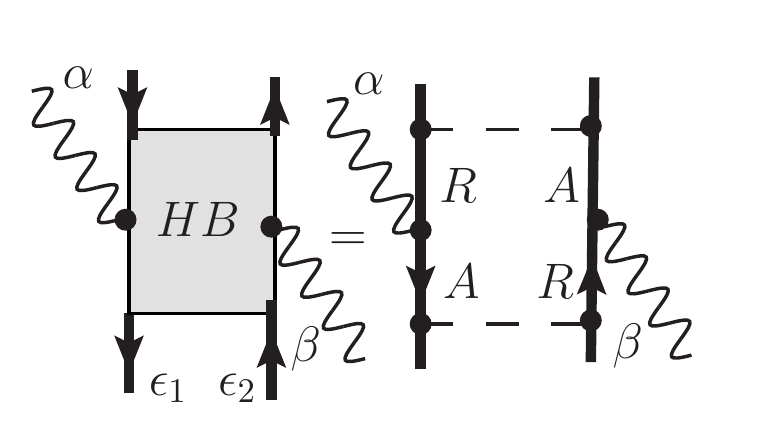}\raisebox{25pt}{$\displaystyle  =4\pi \tau^2 \nu \mathcal{D} \delta_{\alpha\beta}$}}\newline
\makebox[\columnwidth][l]{\raisebox{50pt}{d)}\includegraphics[width=160pt]{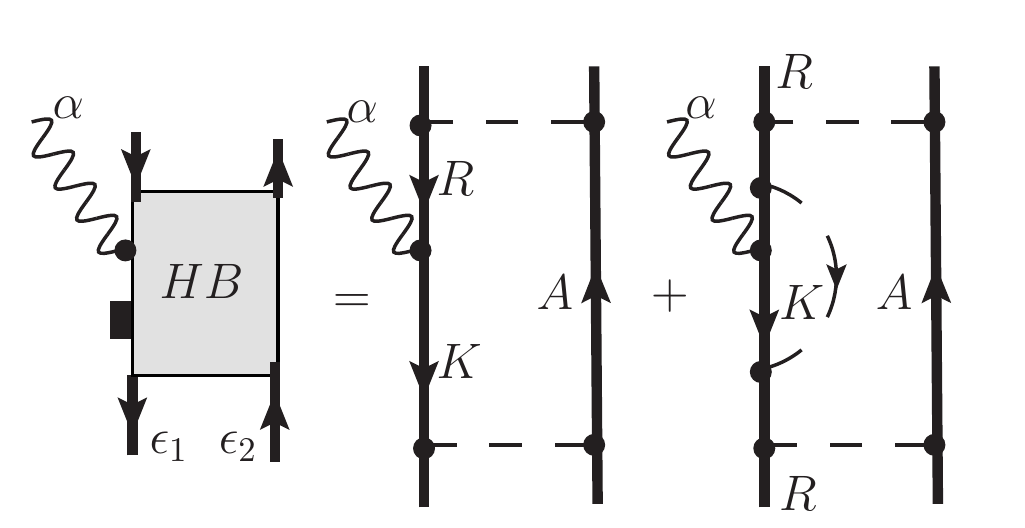}}\newline
\makebox[\columnwidth][l]{\hspace{60pt} $\displaystyle  = 2\pi i\tau^2\ll\!\! j_\alpha\!\!\gg f'_F(\epsilon_2)$}\newline
\makebox[\columnwidth][l]{\raisebox{50pt}{e)}\hspace{13pt}\includegraphics[width=160pt]{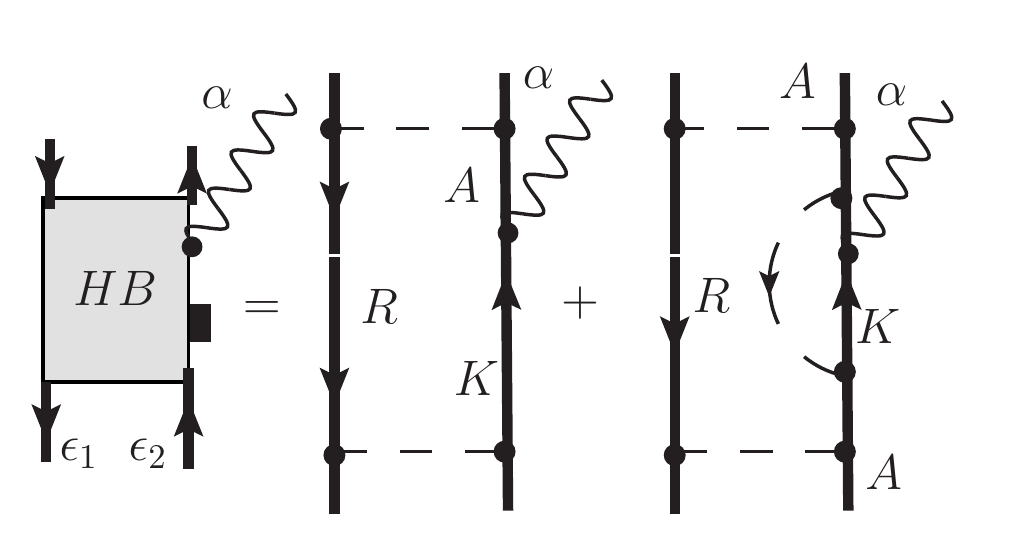}} \newline
\makebox[\columnwidth][l]{\hspace{60pt} $\displaystyle  =2\pi i\tau^2 \ll\!\! j_\alpha\!\!\gg f'_F(\epsilon_2)$}\newline
\caption{\label{fig:Hikami} The Hikami box subdiagrams. The external lines are amputated.}
\end{figure}

In addition to affecting the long range correlations as encoded in the Diffuson and Cooperon, the disorder also affects the short range correlations of operators. This is encoded in the Hikami box subdiagrams shown in Fig. \ref{fig:Hikami}. 

\begin{figure}
\Large
\makebox[\columnwidth][l]{\raisebox{50pt}{a)}\includegraphics[width = 220pt]{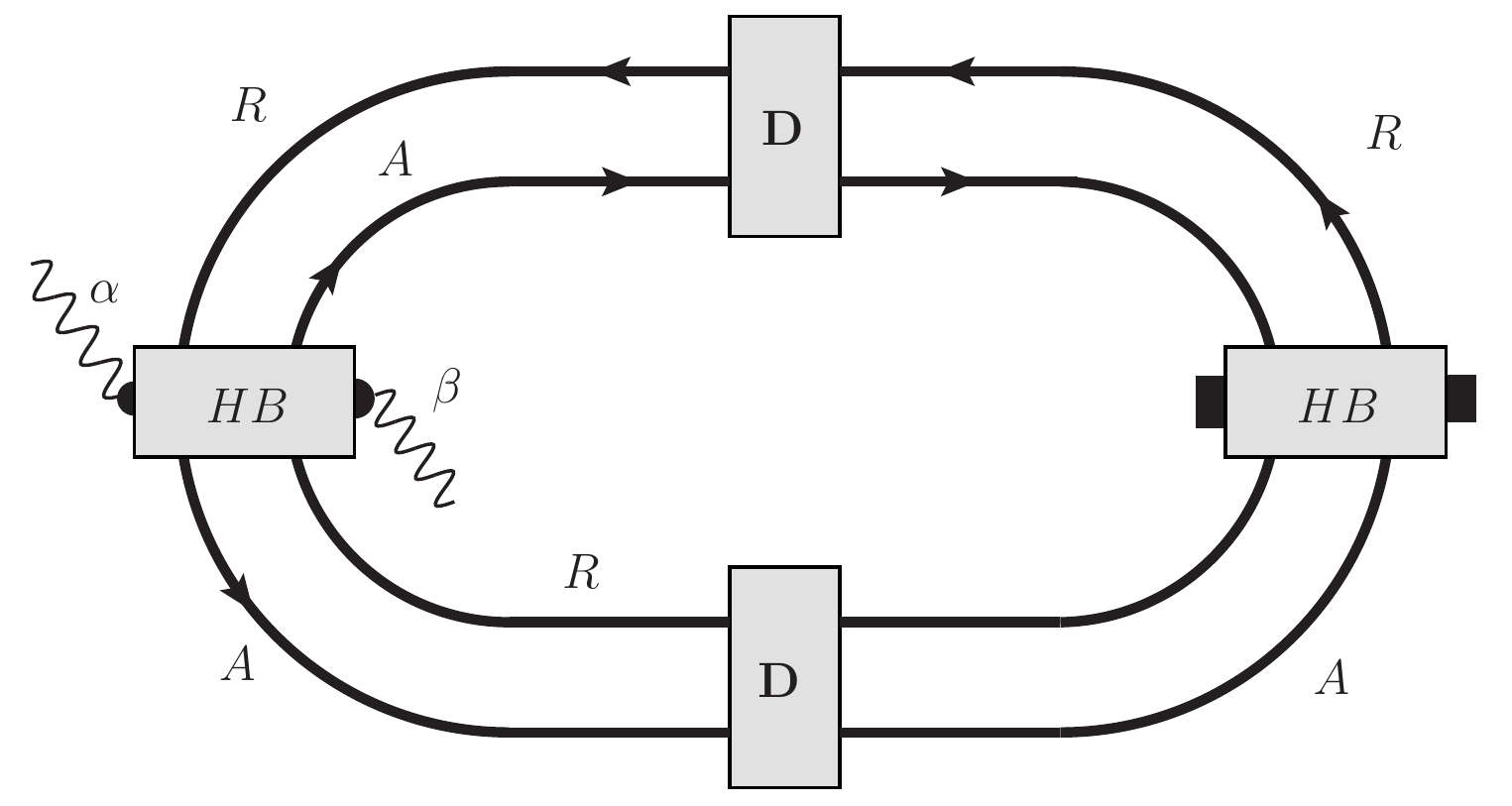}}\newline
\makebox[\columnwidth][l]{\raisebox{50pt}{b)}\includegraphics[width = 220pt]{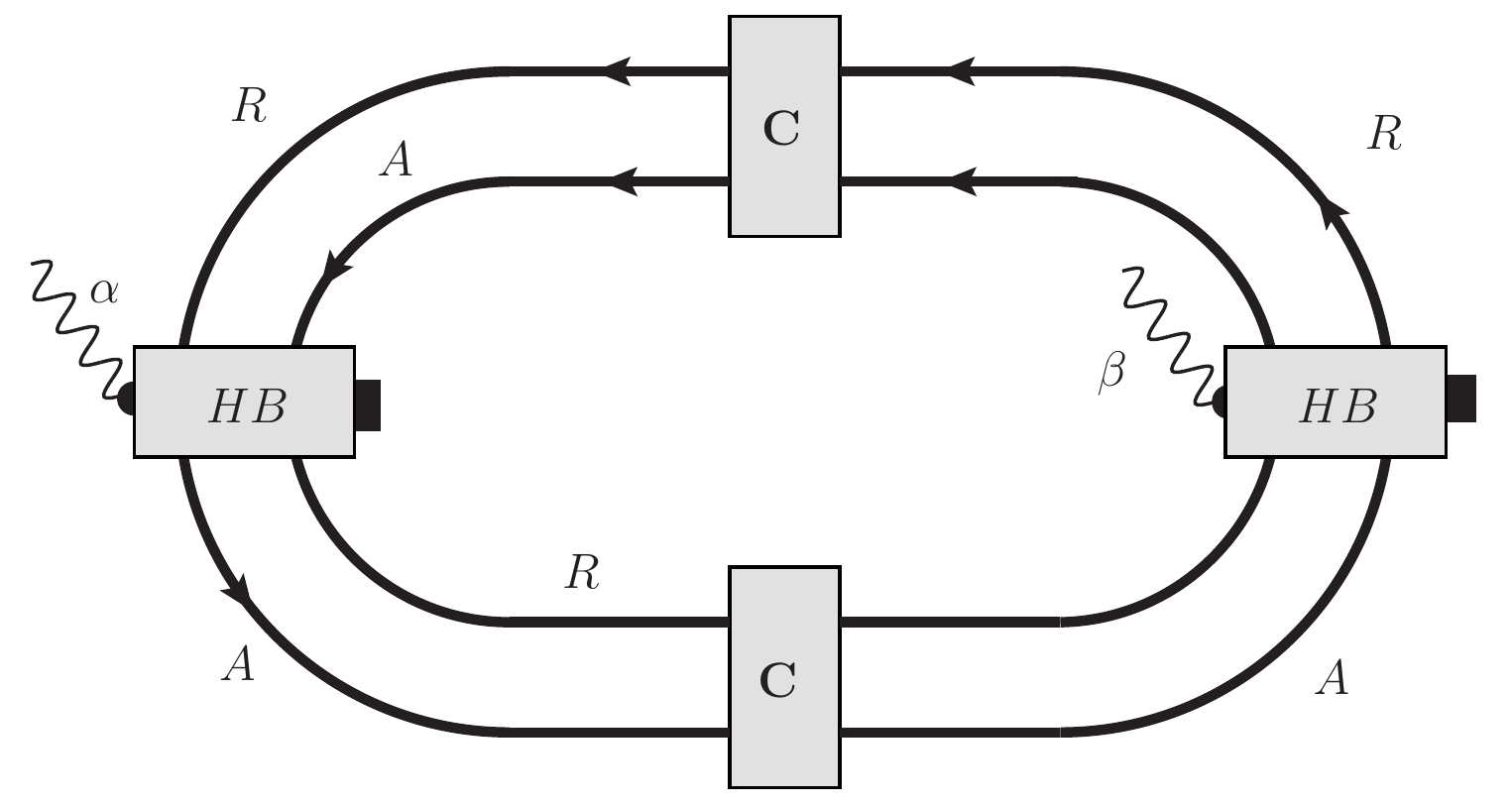}}
\caption{\label{fig:UCF} The diagrams contributing to the universal conductance fluctuations. They must be multiplied by the factor $(-i)^2$. Compare with Fig.~\ref{fig:DistDefs}(d)}
\end{figure}

The mesoscopic fluctuations originate in the dependence of $G^K$ on the disorder. The variance is calculated diagramatically in Fig. \ref{fig:UCF}. In the limit~$T\tau_\phi \gg 1$ calculation yields,
\begin{widetext}\begin{equation} \begin{aligned}
\ll\delta j^\alpha(r,t_1)\delta j^\beta(r',t_2) \gg\, =\left(\pi\nu\right)^{-2}\!\!\int\!\!d\epsilon_1d\epsilon_2\left(\!\frac{\partial f_F}{\partial \epsilon_1}\right)\left(\!\frac{\partial f_F}{\partial \epsilon_2}\right)\Bigg\{ |&\mathbf{C}(\epsilon_1-\epsilon_2,r,r')|^2 j^{\alpha}(r,t_2)j^{\beta}(r', t_1) \\ +\,\,\delta^{(d)}\!(r-r')\delta^{\alpha\beta}\sum_\gamma\! \int\!\! dr'' \bigg[|&\mathbf{D}(\epsilon_1-\epsilon_2,r,r'')|^2  j^\gamma(r'',t_1)j^\gamma(r'',t_2)\bigg]\Bigg\}\label{eq:Noise}
\end{aligned}\end{equation}
\end{widetext}

We now simplify \niceref{eq:Noise}, working in $d<2$ and and analytically continuing. Using the fact $\eta \equiv \epsilon_1 -\epsilon_2$ is of the order of $\tau_\phi$ whereas $\epsilon_{1,2}\sim T$, we may take one of the integrals over $\epsilon$. Further the function $\mathbf{C}(r,r')$ falls off exponentially for $|r-r'|\gg\ell_\phi$. Assuming that $j(r)$ is smooth on the scale $\ell_\phi$, we can remove $j(r)$ from any integral over position. Lastly using the fact that,
\begin{equation}
\begin{aligned}
\int \!\!dr dr' \int_{-\infty}^{\infty}\!\!d\eta\,&|\mathbf{C}\!\left(\eta,r,r'\right)|^2 =\pi\!\int\!\!dr\,\mathbf{C}\left(0;r,r\right)
\end{aligned}
\end{equation} 
we obtain 

\begin{equation} \begin{aligned}
&\ll\!\!\delta j^\alpha(r,t_1)\delta j^\beta(r',t_2)\!\!\gg =\!\delta^{(d)}\!\left(r\!-\!r'\right)\! \sum_{\rho\sigma}\!j^\rho(r,t_1)j^\sigma\!(r',t_2)\\&\qquad \times\frac{1}{3\pi T\nu^2}\Big[\delta^{\alpha\beta}\delta^{\rho\sigma}\mathbf{D}\left(0,r,r\right) +\delta^{\alpha\rho}\delta^{\sigma\beta}\mathbf{C}\left(0,r,r\right)\Big].\label{eq:NoiseReduced}
\end{aligned}
\end{equation}
We now apply \niceref{eq:Noise} to the experimental setup of interest. Consider a cubical system of linear dimension $L$, with leads welded on to the faces normal to the $\hat{x}$ direction. Apply a voltage $V$ and measure the current $I$. To relate $I$ to the local fluctuation $\delta j$ we should recall that the correct interpretation of the term $\delta j(r,t)$ is as a Langevin source for the current density $j(r)$,
\begin{equation}
j^\alpha(r,t) =\sigma E^\alpha + \delta j^\alpha(r,t),
\end{equation}
where $E$ is the electric field and $\delta j(r)$ is to be treated as a random term with statistics given by \niceref{eq:Noise}. However, since we are dealing with a good conductor there is no local charge accumulation on the time scales of interest, as the electric field $E$ compensates instantly. The only effect of the Langevin force $\delta j(r)$ is to affect the charge transport across the system, so the correction to the current $\delta I(t)$,
\begin{equation}
\delta I(t) = \frac{1}{L}\!\int\!\! d^dr\,\, \delta \vec{j}(r,t)\cdot\hat{x}.
\end{equation}
To first order, the current density that appears on the right hand side of  \niceref{eq:NoiseReduced} can be taken to be the Drude result $j = V\sigma L^{2-d} =I L^{1-d}$ giving [compare with \niceref{eq:UCFReal}],
\begin{equation}
\begin{aligned}
\ll \delta I(t)\delta I(0)\gg\, =\,& \frac{I^2}{L^d}\frac{\ell_\phi^d}{T\tau_\phi g(\ell_\phi)^2} \bigg[Y\!\left(\!\frac{\ell_\phi}{\ell_{B_-}}\!\right) + Y\!\left(\!\frac{\ell_\phi}{\ell_{B_+}}\!\right)\!\bigg]\\&+ \frac{I^2}{L^d}\frac{L^d_T}{g\left(L_T\right)^2}f_d, \label{eq:NoiseReduced2}
\end{aligned}
\end{equation}
where $L_{B_\pm} \equiv \left(e\left|B(0)\pm B(t)\right|\right)^{-1/2}$ and $Y$ is the scaling function defined by 
\begin{equation}Y\left(\frac{\ell_\phi}{\ell_B}\right) \equiv \frac{1}{3\pi}\frac{\ell_\phi^d}{\tau_\phi} \left[\mathbf{C}\left(0,r,r;B\right)-\mathbf{C}\left(0,r,r;0\right)\right]. \end{equation}
This function is well known from the study of weak localization and see Refs.~[\onlinecite{ALK, HLN}] for evaluation. The magnetic field indepedent term $f_d$ appears on analytic continuation to $d=2,3$. In $d=2$ it is given by,
\begin{equation}
f_2  = \frac{1}{6\pi^2}\log\left(\frac{1}{T\tau}\right),
\end{equation}
and is a nonuniversal constant in $d=3$.
 
\subsection{$1/f$ noise\label{sec:1f}}
\begin{figure}
\large
\makebox[\columnwidth][l]{\raisebox{25pt}{a)}\includegraphics[width=45pt]{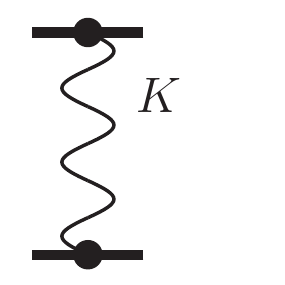}\hspace{-10pt}\raisebox{15pt}{$=\displaystyle\frac{1}{2\pi\nu\tau_*}F^K\!\!\left(t_1-t_2\right)$}}\newline
\makebox[\columnwidth][l]{\raisebox{20pt}{b)}\includegraphics[width=220pt]{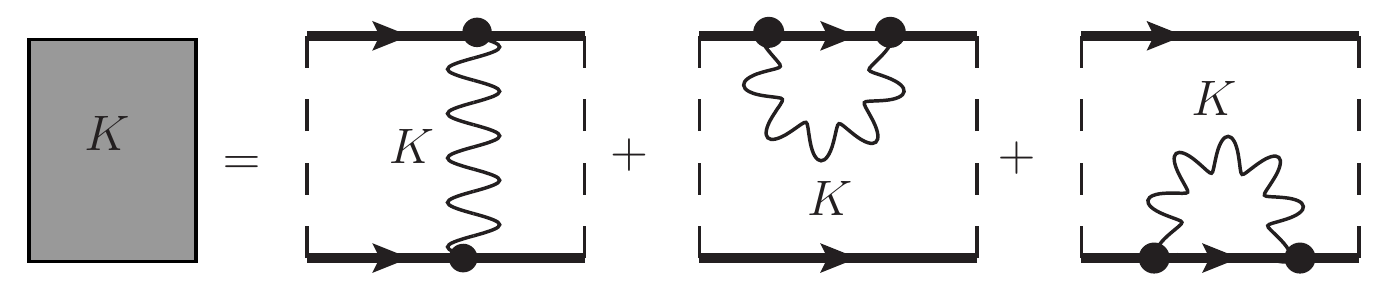}}\newline
\makebox[\columnwidth][l]{\raisebox{20pt}{\hspace{50pt}$=\displaystyle\frac{2\pi\nu\tau^2}{\tau_*}\left[F^K\!\left(t_1-t_2\right)-F^K\!\left(0\right)\right]$}}\newline
\makebox[\columnwidth][l]{\raisebox{100pt}{c)}\includegraphics[width=220pt]{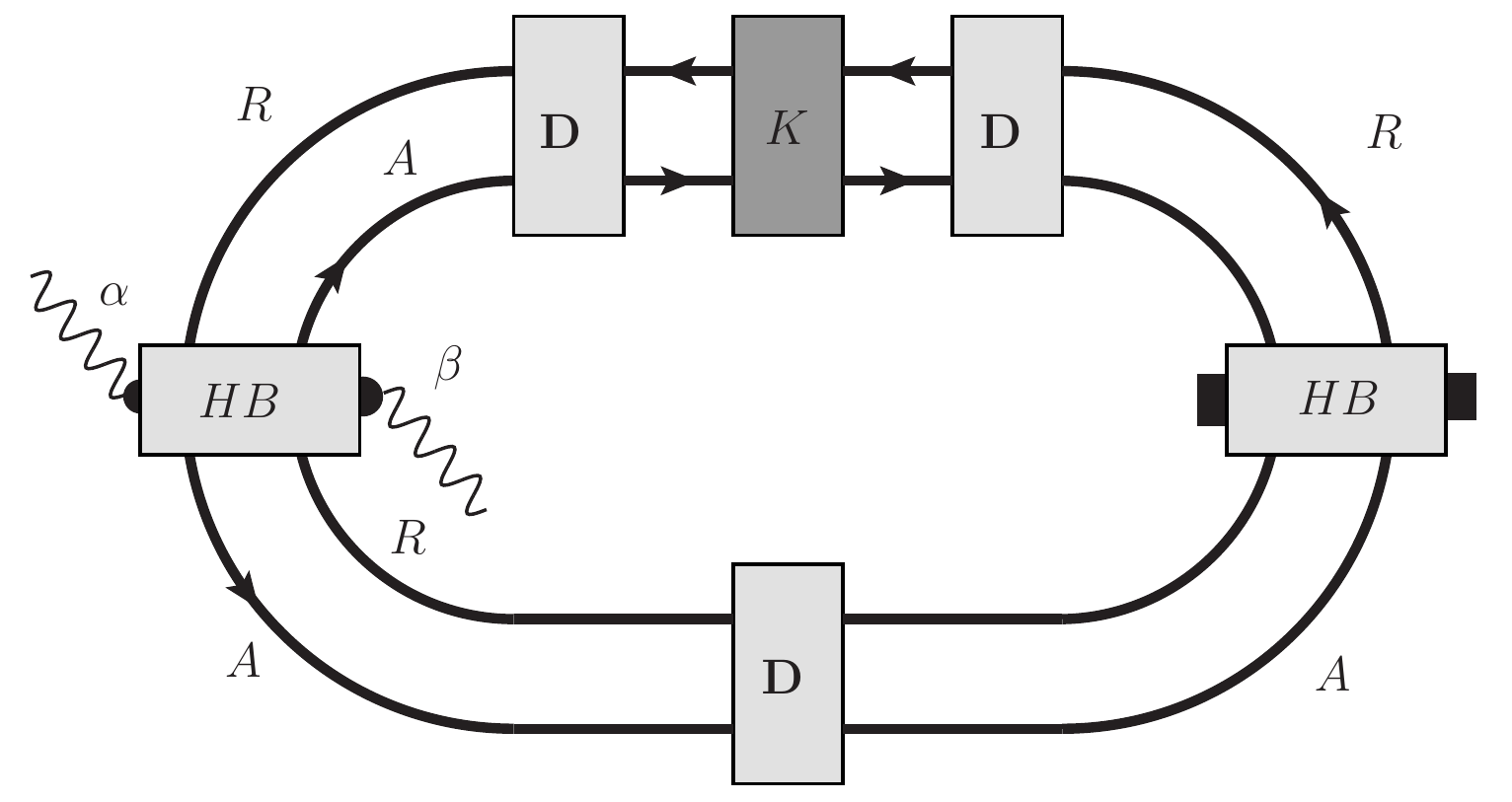}}\newline
\makebox[\columnwidth][l]{\raisebox{100pt}{d)}\includegraphics[width=220pt]{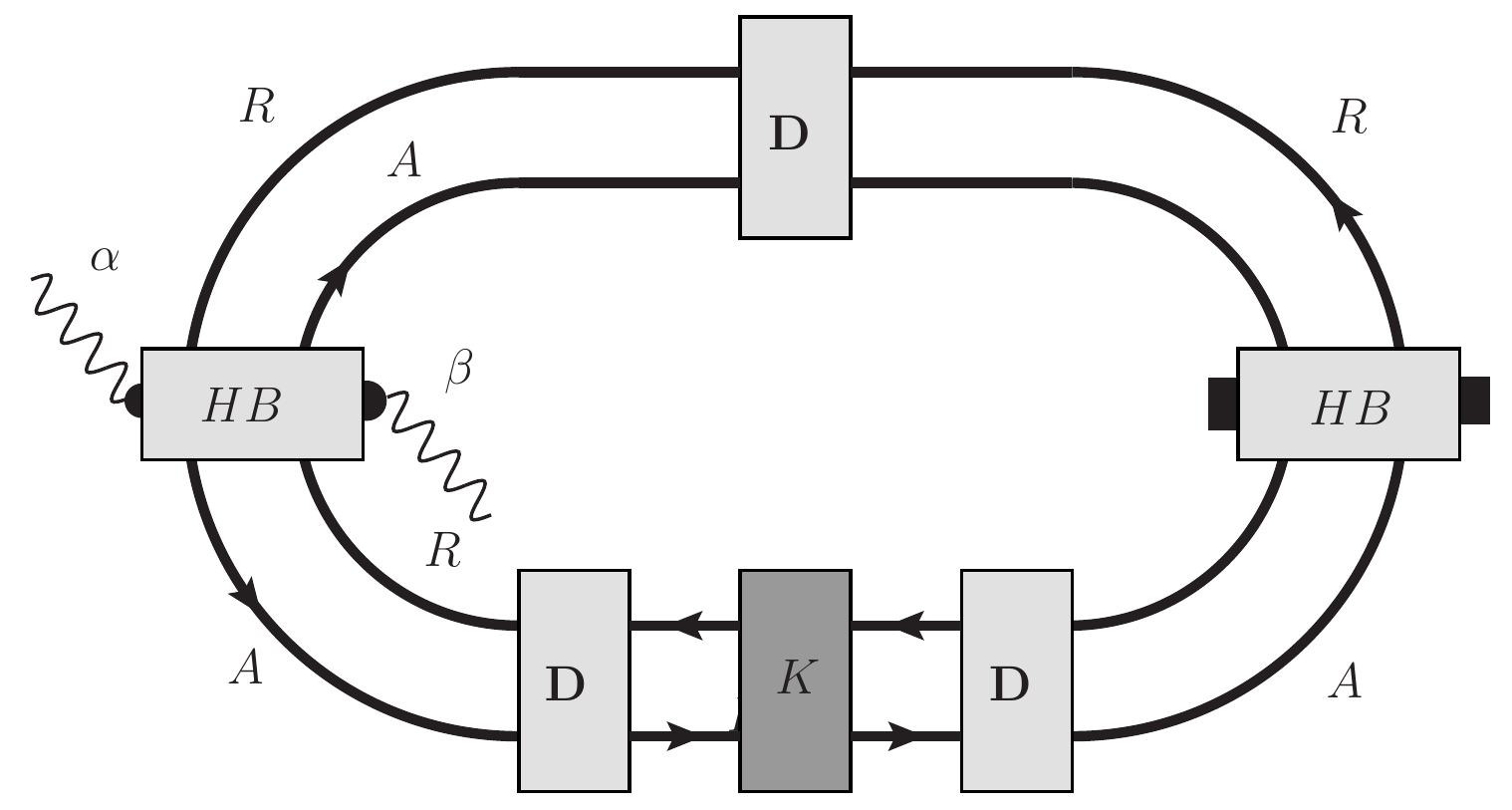}}\newline
\makebox[\columnwidth][l]{\raisebox{100pt}{e)}\includegraphics[width=220pt]{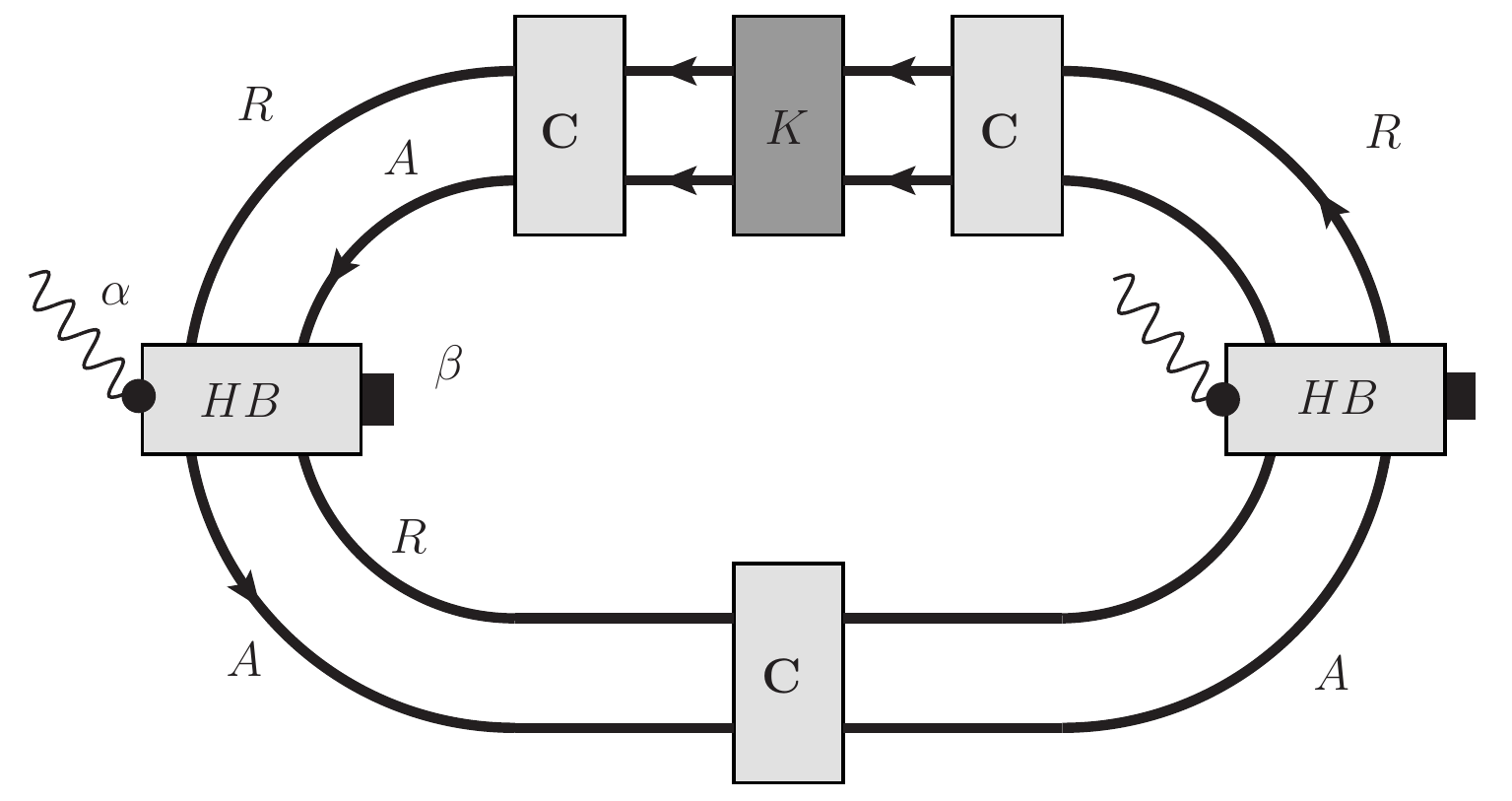}}\newline
\makebox[\columnwidth][l]{\raisebox{100pt}{f)}\includegraphics[width=220pt]{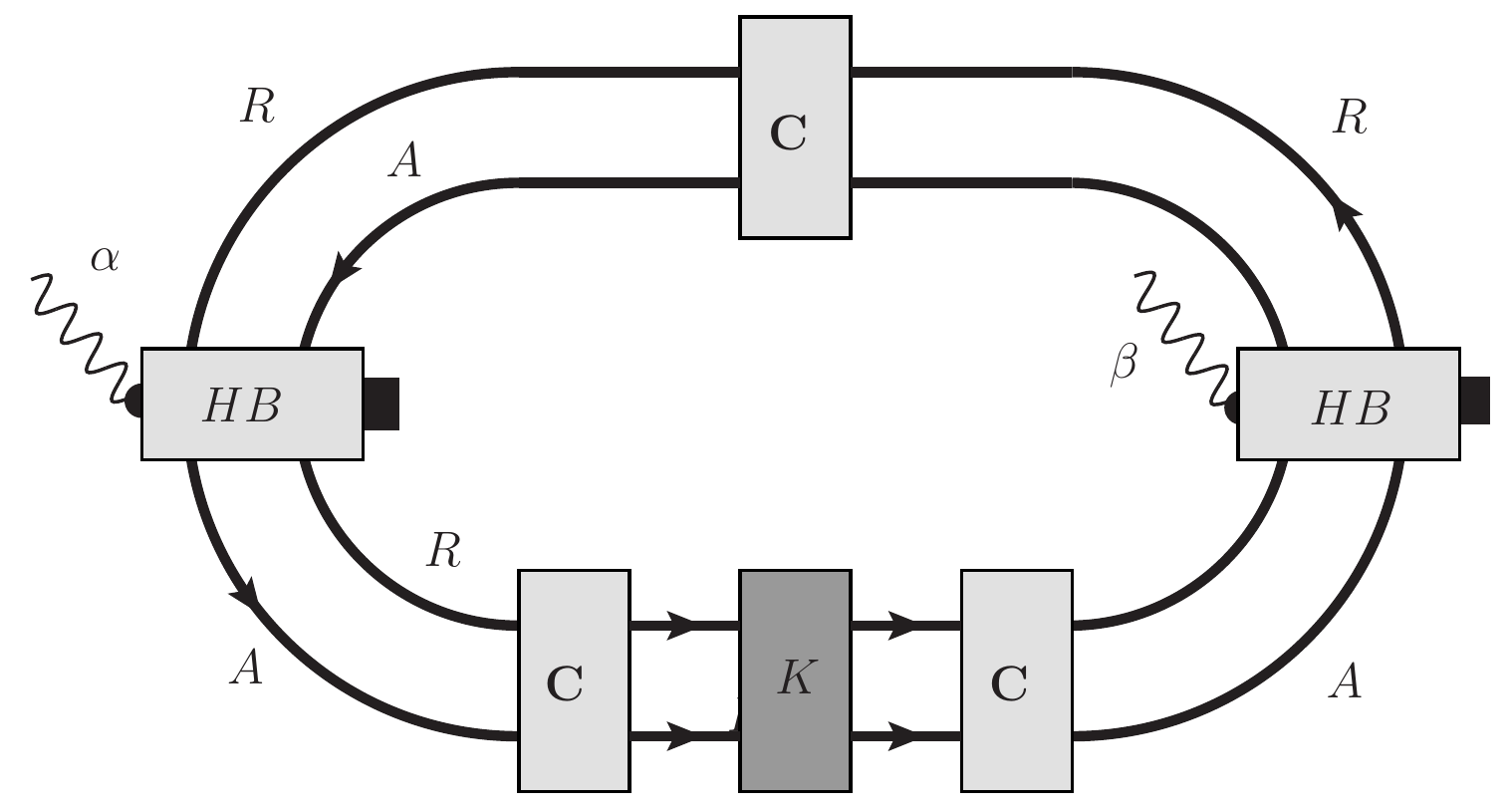}}\newline

\caption{\label{fig:Noise} The diagrams contributing to the noise. The TLS enter through subdiagram (b). }
\end{figure}

The mesoscopic fluctuations can be made observable by varying an external parameter, such as magnetic field. The shifting of the TLS is another mechanism by which the mesoscopic fluctuations are manifested, in this case as the $1/f$ noise. The appropriate diagrams are collected in Fig. \ref{fig:Noise}.
 In fact, no new calculation is needed since we may use the result for the mesoscopic fluctuation (\ref{eq:Noise}), make the substitution $\tau_\phi^{-1}\rightarrow \tau_\phi^{-1} + \tau_*^{-1}\left(\bar{F}^K(t) - \bar{F}^K(0)\right)$ and then expand to first order. The resulting correlations of the current are
\begin{equation} \begin{aligned}
&\ll\!\!\delta j^\alpha(r,t_1)\delta j^\beta(r',t_2)\!\!\gg =\!\delta^{(d)}\!\left(r\!-\!r'\right)\! \sum_{\rho\sigma}\!j^\rho(r,t_1)j^\sigma\!(r',t_2)\\& \times\frac{1}{3\pi\nu^2T\tau_*}\frac{\partial}{\partial \tau_\phi^{-1}}\Big[\delta^{\alpha\beta}\delta^{\rho\sigma}\mathbf{D}\left(0,r,r\right) +\delta^{\alpha\rho}\delta^{\sigma\beta}\mathbf{C}\left(0,r,r\right)\Big].\label{eq:NoiseSimplified}
\end{aligned}
\end{equation} We may follow the same arguments as above to translate this expression into an expression for the fluctuations of the current. In terms of the function $\mathcal{F}$ (see \niceref{eq:DefCalF}),

\begin{equation}\mathcal{F}(t) =L^d(\ll\delta I(0)\delta I(t)\gg)/I^2,\label{eq:FAutoCorr}\end{equation}

the result is 
\begin{equation}
\begin{aligned}
\mathcal{F}(t) &= \frac{\ell_\phi^d}{T\tau_* g(\ell_{\phi})^2 }\left[Z \left(\frac{\ell_\phi}{\ell_{B_-}}\right) +Z\left(\frac{\ell_\phi}{\ell_{B_+}}\right)\right]\\&\times\left[F^K(t) - F^K(0)\right],
\end{aligned}
\label{eq:CalcCalF}
\end{equation}
where,
\begin{equation}
Z(x) = \left(d/2 -1\right) Y(x) - 2xY'(x) +\frac{1}{12\pi^2}\delta_{2,d}.\end{equation}
The final term of \niceref{eq:CalcCalF}, in square brackets, carries all of the details of the microscopic model.  The noise can therefore be used calculate $\tau_*$ and the correlations of the impurities.

On insertion of the result for the TLS (See Appendix A) becomes 
\begin{equation}
\mathcal{F}(t) \propto  -\frac{\log(t/t_0)}{\log(t_m/t_0)},\end{equation}
for times $t$ with $t_0 < t  < t_m$. For frequencies $f$ with $ t_0 < f^{-1} < t_m$ the Fourier transform of the autocorrelation has the expected $1/f$ scaling. Given that $t_0$ is microscopic while $t_m$ may be on the order of a day, this reproduces the experimental fact of $1/f$ scaling over many orders magnitude.

\subsection{Memory effect\label{sec:MemEffect}}

We now calculate the memory effect, which is the correction to the conductivity arising from the past history of the chemical potential $\mu(t)$ and magnetic field $B(t)$. 
By quickly sweeping the chemical potential at well separated times, the entire time history of the conductivity at all energies may be reconstructed. Throughout this section we will suppress the dependence of $\mathbf{C}$ and $\mathbf{D}$ on magnetic field.

\begin{figure}
\large
\vspace{-9pt}
\makebox[\columnwidth][l]{\raisebox{25pt}{a)}\includegraphics[width=45pt]{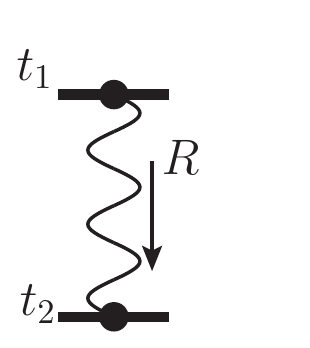}\hspace{-10pt}\raisebox{20pt}{$=\displaystyle\frac{i}{2\pi\nu\tau_*}F^R\!\!\left(t_1-t_2\right)$}}\newline
\makebox[\columnwidth][l]{\raisebox{20pt}{b)}\includegraphics[width=100pt]{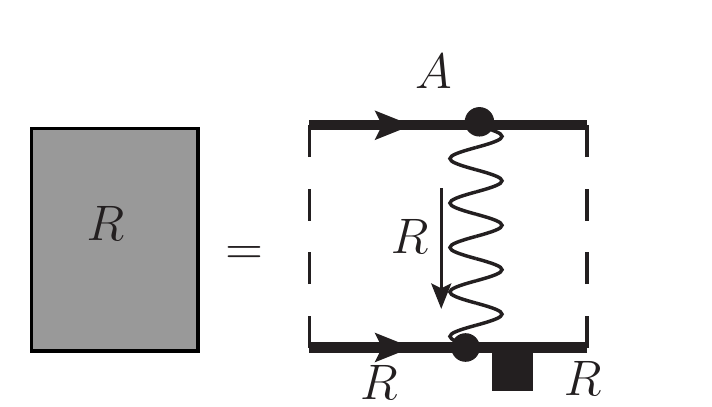}}\newline
\vspace{-11pt}
\makebox[\columnwidth][l]{\hspace{50pt}$\!\!=\!\displaystyle\frac{2\pi i\nu\tau^2}{\tau_*}F^R\!\left(t_1-t_2\right)f_F\!\left(\epsilon-\mu(t_2)\right)$}\newline
\vspace{-6pt}
\makebox[\columnwidth][l]{\raisebox{100pt}{c)}\includegraphics[width=215pt]{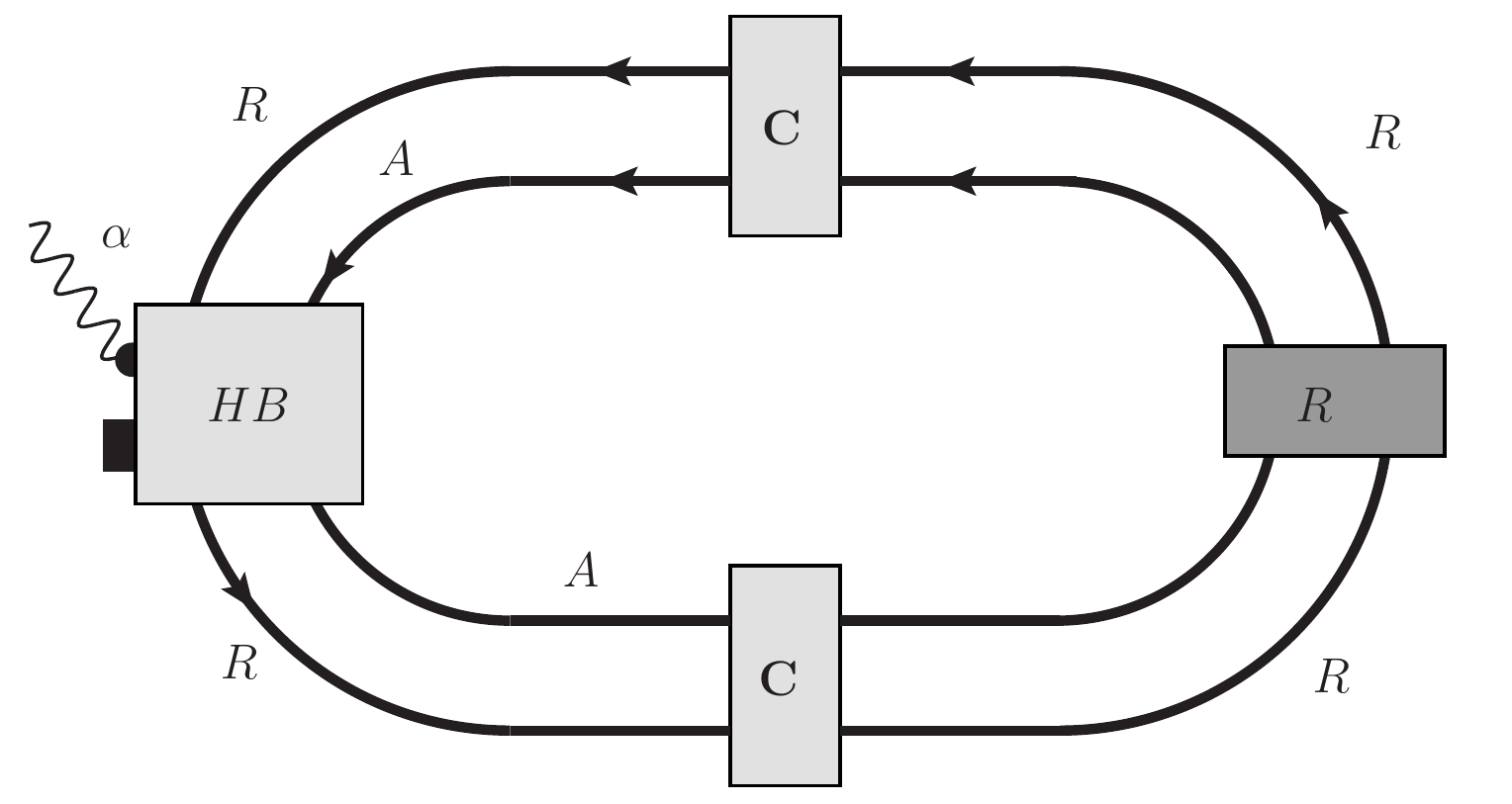}}\newline
\makebox[\columnwidth][l]{\raisebox{100pt}{d)}\includegraphics[width=215pt]{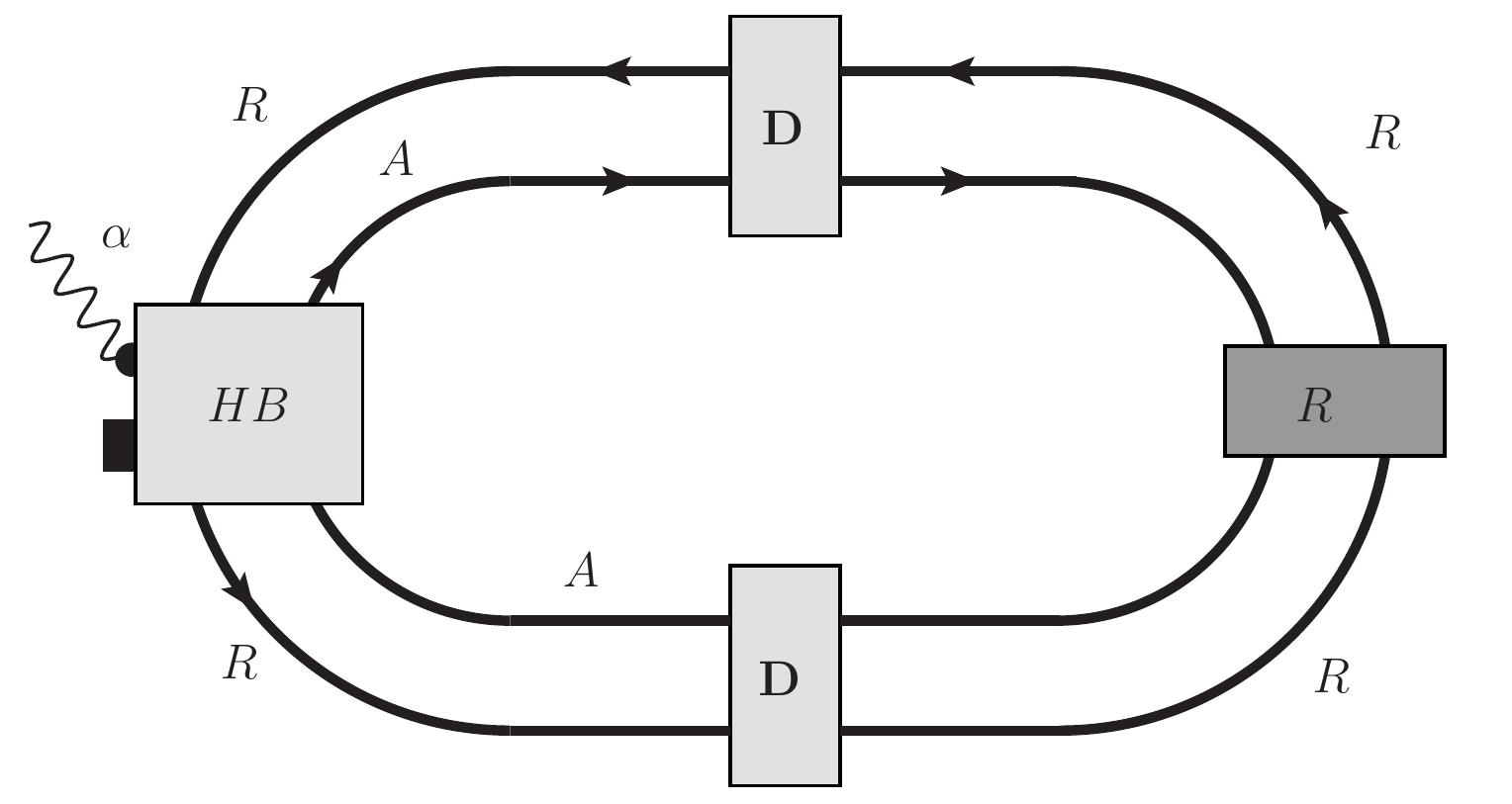}}\newline
\makebox[\columnwidth][l]{\raisebox{100pt}{e)}\includegraphics[width=215pt]{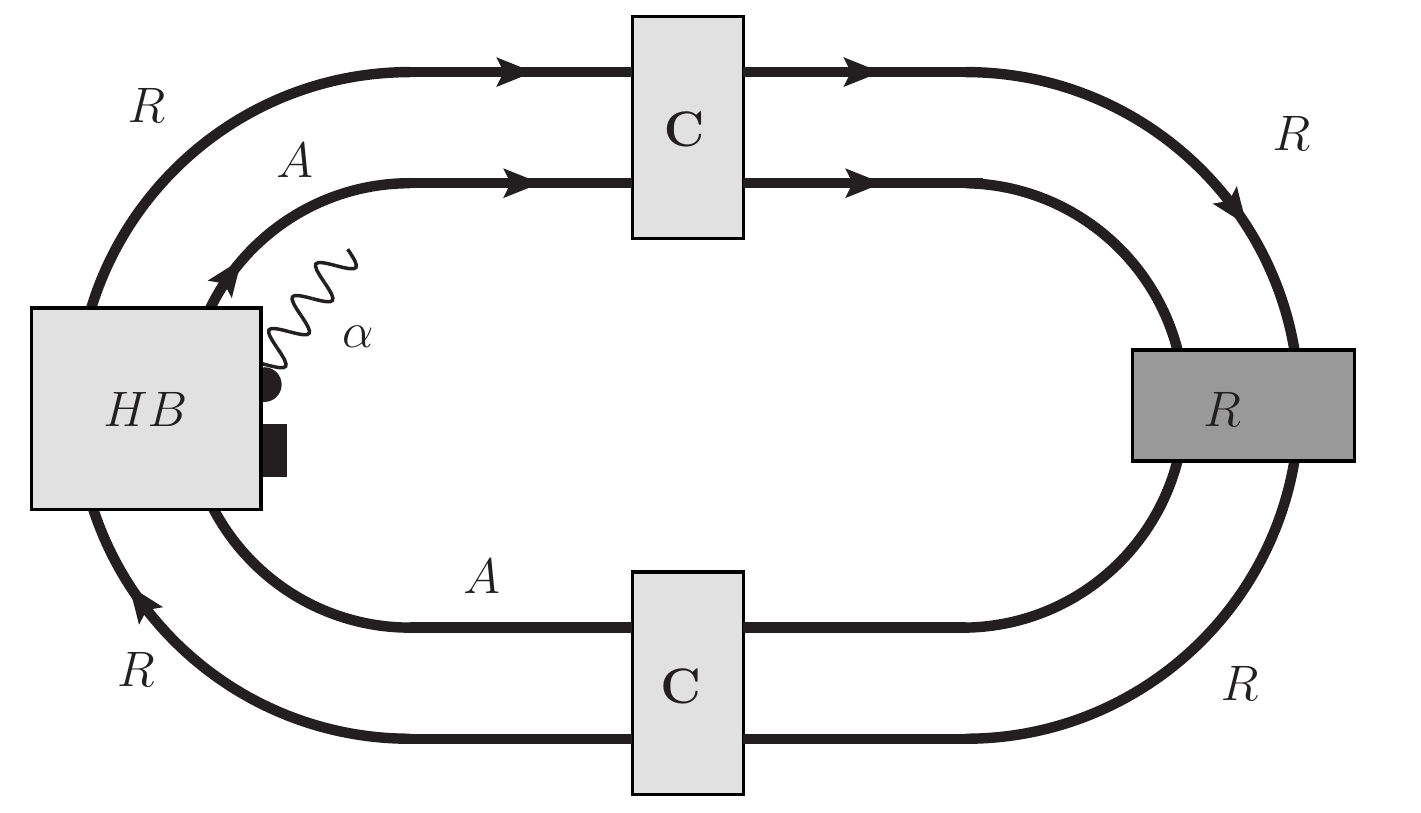}}\newline
\makebox[\columnwidth][l]{\raisebox{100pt}{f)}\includegraphics[width=215pt]{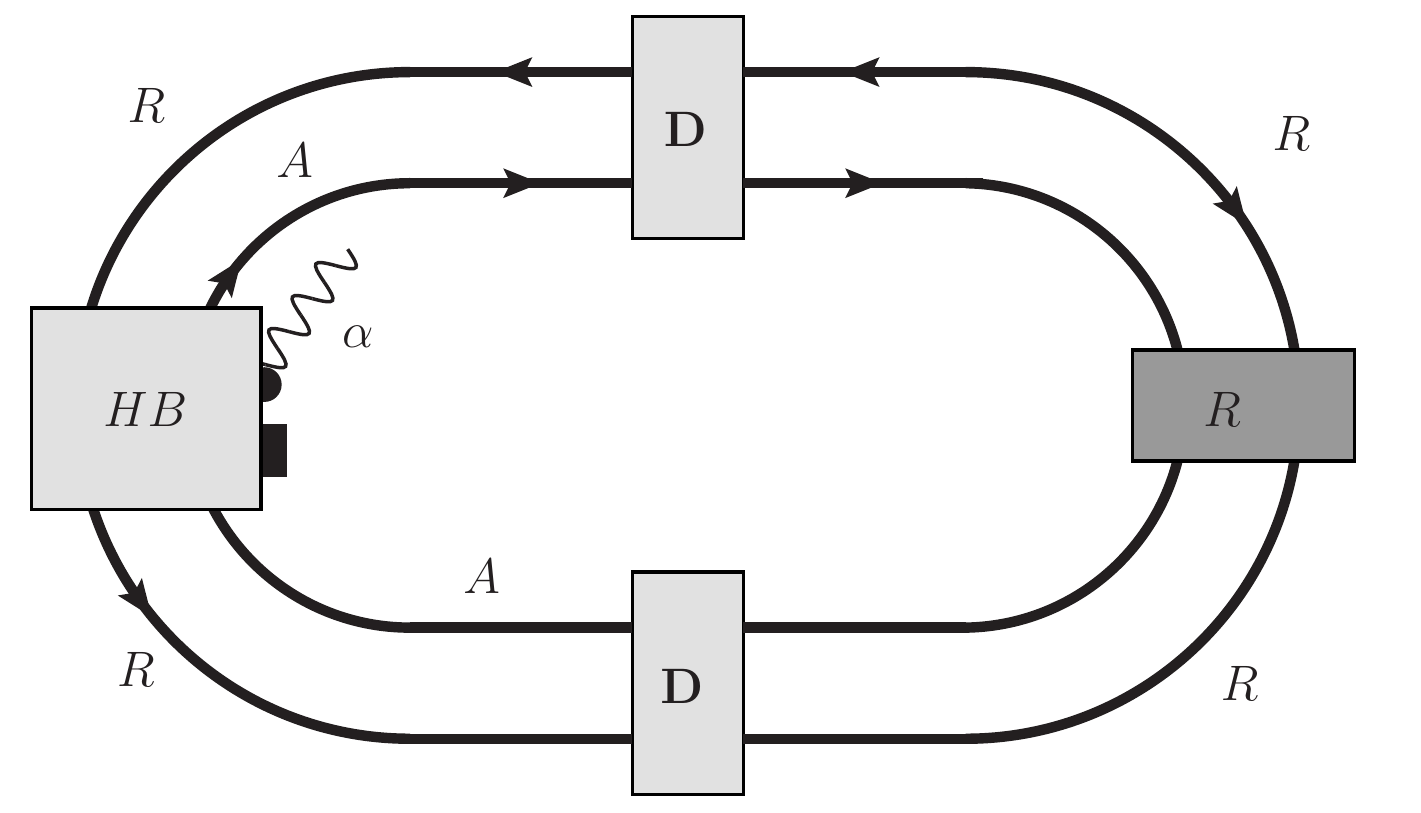}}\newline
\caption{\label{fig:MemEffect} The diagrams contributing to the memory effect. The TLS enter through subdiagram (a). Note there is an overall factor of $i$ from the definition of $j$ in Fig. \ref{fig:DistDefs}(c) }
\end{figure}

 The corrections to the measured current are shown in Fig.~\ref{fig:MemEffect}.
The history of the system parameters $\mu(t)$ and $B(t)$ enter through the history of the electron occupation function, $n_{\epsilon}(t) = \tanh\left(\frac{\epsilon - \mu(t)}{2T}\right)$. The correction to the measured conductivity is,
\begin{equation}
\begin{aligned}
\frac{\delta\sigma(t)}{\sigma_D} =& \int \!d\,t'\frac{F^R(t-t')}{\nu\tau^*}\\&\times\int\!\frac{d \epsilon}{2\pi}\frac{d \epsilon'}{2\pi}X(\epsilon'- \epsilon)\frac{\partial n_\epsilon(t)}{\partial \epsilon} n_{\epsilon'}(t') .
\end{aligned}
\end{equation}
It is important to note the the energies in the distribution function are defined relative to the chemical potential at the time $t$. The kernel $X$ is defined by
\begin{equation}
X(\eta) =2\mathfrak{Re}\bigg\{i\!\! \int\!\!\frac{d^d\vec{Q}}{(2\pi)^d}\left[\mathbf{C}(\vec{Q},\eta)^2 + \mathbf{D}(\vec{Q},\eta)^2\right]\bigg\}.
\end{equation}
The integral over $\eta$ and $Q$ is not convergent in $d=2$ and $d=3$, so there are logarithmic term in $d=2$ and non-universal constant terms in $d=3$.  Using the fact that $\mathcal{C}(\eta)^2 =-i\partial_\eta\mathcal{C}(\eta)$ and likewise for the difuson, we can integrate by parts, obtaining:
\begin{equation}\begin{aligned}
\int_{-\infty}^\infty\!\!& \frac{d \epsilon'}{2\pi} X(\epsilon'- \epsilon)n_{\epsilon'}(t') = \\ \mathfrak{Re}\bigg\{& \int_{-\infty}^\infty \frac{d \epsilon'}{2\pi}\frac{\partial n_{\epsilon'}(t')}{\partial\epsilon'}\left[\mathbf{C}(r,r,\epsilon'-\epsilon)\ +\mathbf{D}(r,r,\epsilon'-\epsilon) \right]\bigg\}.\end{aligned}
\end{equation}

Finally, using the fluctuation dissipation relationship between $F^K$ and $F^R$  [see \niceref{eq:FlucDiss}], we obtain the main result of this section: 
\begin{equation}
\begin{aligned}
\frac{\delta\sigma(t)}{\sigma_D} &=\frac{1}{T\tau_*}\frac{1}{g\left(L_T\right)} \int_{-\infty}^{t}\!\!dt'\,\, \frac{d\bar{F}^K(t-t')}{d t}\bigg[\\&\!\!\!\!S\left(\frac{\mu(t) - \mu(t')}{T}, \frac{L_T}{L_{B_+}}\right) +S\left(\frac{\mu(t) - \mu(t')}{T}, \frac{L_T}{L_{B_-}}\right)\bigg],
\end{aligned}
\label{eq:MemEffectMain}
\end{equation}
compare with \niceref{eq:ZBA}. The conducance at scale T is determined by the scaling
\begin{equation}\frac{g\left(L_T\right)}{g\left(\ell_\phi\right)} \equiv\left(\frac{L_T}{\ell_\phi}\right)^{2-d}, \end{equation} 
and the magnetic length $L_{B_{\pm}}$ is defined by
\begin{equation}
L_{B_\pm} \equiv \sqrt{\frac{\hbar c}{e\left|B(t)\pm B(t')\right|}}.
\end{equation}
The scaling \footnote{There also may be an effect of the magnetic through the Zeeman coupling, but this should be a secondary effect.} function $S$ is defined by,
\begin{equation}
S\left(u,v\right)\equiv 8\int_{-\infty}^{\infty}\!\!dx\,\,\frac{x\text{coth}x -1 }{\text{sinh}^2x}\mathfrak{Re}\bigg\{\bar{\mathbf{C}}\left(0,(2x + u), v\right)\bigg\}.
\label{eq:MemEffectScaling}\end{equation}
Here $\bar{\mathbf{C}}$ is the Cooperon expressed in dimensionless units, given by the equation,
\begin{equation}
\left[iu + \left(i\vec{\nabla} +  v\bar{A}(r)\right)^2\right]\bar{\mathbf{C}}\left(r, u, v\right) =\delta^{(d)}(r),
\end{equation}  
where $\tilde{A}$ is a dimensionless gauge potential obeying,
\begin{equation}
\vec{\nabla}\times \tilde{A} = \hat{z},
\end{equation} 
and $\hat{z}$ is the unit vector in the direction of the magnetic field. Although \niceref{eq:MemEffectScaling} only contains the symbol $\bar{\mathbf{C}}$, it includes the Diffuson contribution through the second term of \niceref{eq:MemEffectMain}.  The correction is similar to the usual quantum correction to conductance, but around the old chemical potential.

 The integral over $x \equiv 2(\epsilon_1 - \epsilon_2)/T$ serves to smooth the result over the scale of the temperature. 
At zero magnetic field we may evaluate $S$ explicitly and we obtain
\begin{equation}
S\left(u,0\right) = \int_{-\infty}^{\infty}\!\!dx\,\,\frac{x\,\text{coth}x -1 }{\text{sinh}^2x} P_d\left(2x+u\right).\label{eq:LineShape}
\end{equation}
The function $P_d$ depends on the dimension and is given by
\begin{equation}
\begin{aligned}
&P_1(z) \equiv \frac{2}{\sqrt{2}}\left|z\right|^{-1/2}\\
&P_2(z)\equiv -\frac{2}{\pi} \log\left| \frac{1}{z(T\tau)}\right|\\
&P_3(z)\equiv a - \frac{\sqrt{2}}{\pi}\left|z\right|^{1/2},
\end{aligned}
\end{equation}
where $a$ is a non-universal constant.
When $u\gg1$, $S$ has the limiting form 
\begin{equation}
S(u,0) =P_d(u).
\end{equation}

We now calculate the effect of a transverse magnetic field in $d=2$. In a magnetic field the Cooperon must be expanded in Landau levels,
\begin{equation}
\bar{\mathbf{C}}\left(0,u,v\right)  = \frac{4v^2}{4\pi}\sum^{\infty}_{n=0}\left[i\eta +\left(n+\frac{1}{2}\right)4v^2\right]^{-1}.
\end{equation}
Introducing an integral over the auxiallary variable $s$ this may be rewritten as
\begin{equation}
\begin{aligned}
&\bar{\mathbf{C}}\left(0,u, v\right) - \mathbf{C}\left(0,u,0\right)   =\\&\frac{1}{4\pi} \int_0^{\infty}\!\!\frac{ds}{s}\exp\left(-i\frac{\eta}{T} s\right)\left(\frac{2v^2s}{\text{sinh}\left(2v^2s\right)} - 1\right).
\end{aligned}
\end{equation}
The change in the lineshape $S(u,v)$ can now be evaluated with the result that 
\begin{equation}\begin{aligned}
S(u,v) - S(u,0)&=\frac{2}{\pi}\int_0^{\infty}\frac{ds}{s}\left(\frac{2v^2s}{\text{sinh}\left(2v^2s\right)} - 1\right)\\ &\times\,\,\left(\frac{2\pi s}{\text{sinh}(2\pi s)}\right)^2 \cos(u s).\end{aligned}\end{equation}
Proceeding in the regime where $v \ll 1$, the bulk of the integral comes from the region near zero where the first term may be perturbatively expanded,
\begin{equation}
S(u,v) \approx  S(u,0) + v^4 H(u),\end{equation}where
 \begin{equation}H(u) = \frac{4}{3\pi} \int_0^\infty\!\! dx\, \frac{(2\pi x)^2 x}{\text{sinh}^2(2\pi x)}\cos(ux).\label{eq:RedMagneticCorrection}
\end{equation}

\begin{figure}
\large
\makebox[\columnwidth][l]{\raisebox{25pt}{a)}\includegraphics[width=55pt]{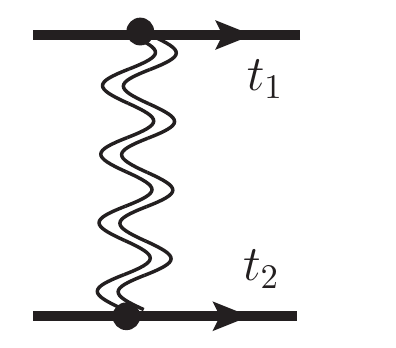}\hspace{-10pt}\raisebox{20pt}{$\displaystyle\propto\delta\left(t_1-t_2\right)$}}\newline
\makebox[\columnwidth][l]{\raisebox{100pt}{b)}\includegraphics[width=220pt]{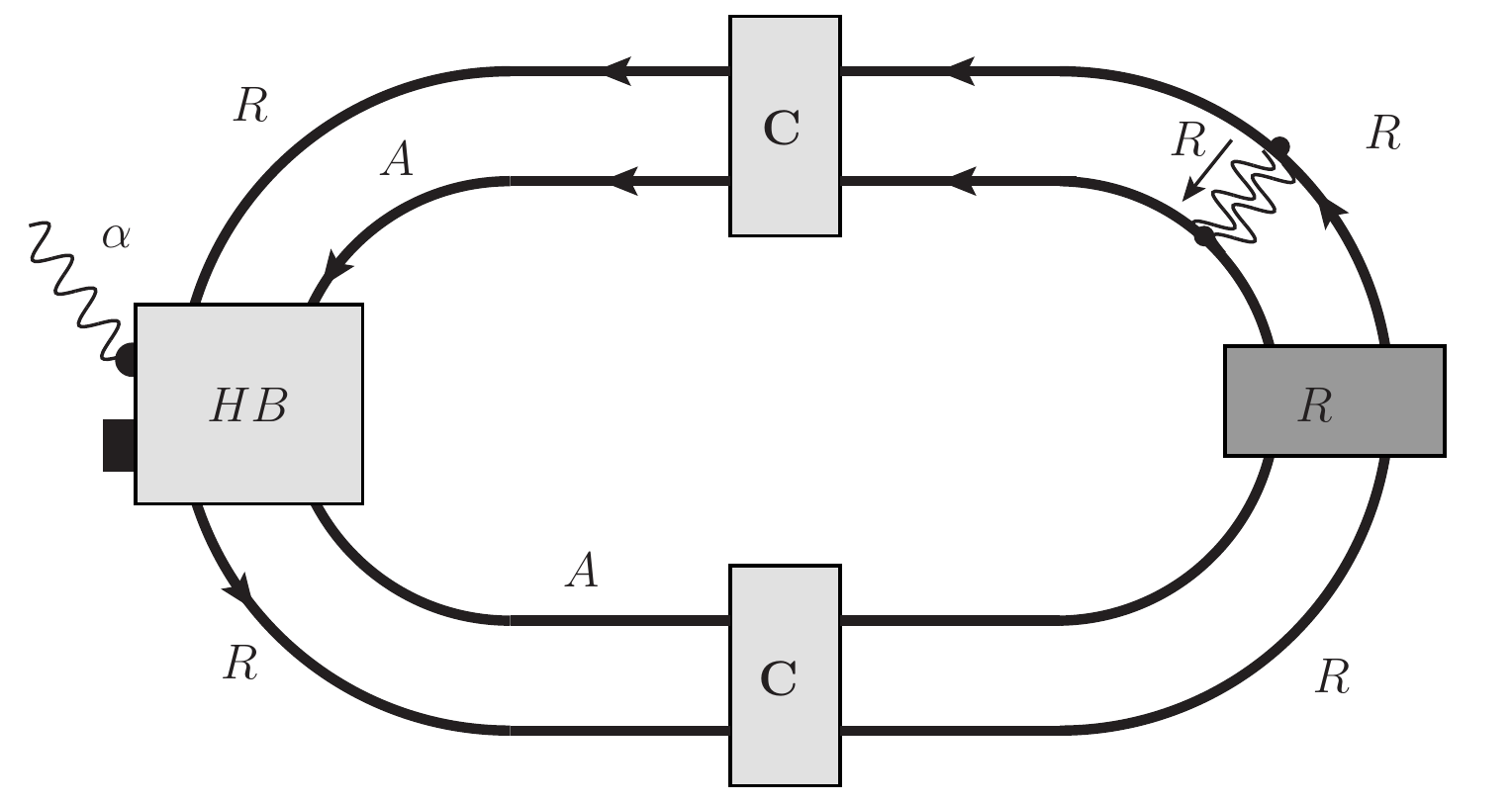}}\newline
\caption{\label{fig:Interaction}  A normal electron-electron interaction, indicated by the double wavy line in figure (a), is effectively a delta function in time on the scales of interest. Therefore diagrams of the form (b) do not contribute to the memory effect and there is no Fermi-liquid type resummation. }
\end{figure}
Finally, although there is a superficial resemblance between the retarded line $F^R$ and the usual electron-electron interactions, the term $F^R$ does not get simply resummed in the usual Fermi-liquid fashion, see Fig.~\ref{fig:Interaction}. This is because any interaction between an electron at time $t_1$ and $t_2$ will produce make the diagram proportional to $\delta\left(t_1 -t_2\right)$ and therefore not contribute to the memory effect.  Reference~[\onlinecite{vonOppen97}] showed that in $d=2$ electron-electron interactions can produce $1/f$ noise, but this is only true for frequencies $f L^2/\mathcal{D}\gg 1$ and thus has no relevance for the longest time behavior in mesoscopic systems.
\section{Conclusion}

The essential conclusions of this paper are as follows: The existence of the two level systems that have been suggested to cause the $1/f$ noise in metals, \emph{necessarily} leads to a memory effect. The strength of the memory effect is universally related to the strength of the $1/f$ noise. 
The lineshape of the memory effect is also a universal function. Since the effects are related to the mesoscopic fluctuations they are sensitive to the magnetic field in a universal fashion. The sensitivity to the Aharonov-Bohm effect, which leads to the magnetic field dependence, is a universal feature of quantum coherent systems.

We emphasize that the conclusions here do not depend on the microscopic model of the TLS. Ghe TLS do not have to be structural defects or mobile impurities.
Any set of localized systems that produce low-frequency noise will, by the fluctuation-dissipation theorem, lead to a long-time memory effect following the universal relationship. 
There is no necessity for the spectrum to be exactly of the form $1/f$ - any slowly decaying spectrum will lead to a memory effect.
Even a mechanism such as atoms diffusing through a network of tunnelling sites - while not in some sense a "localized system" - will still lead to the same relationship between noise and memory\footnote{We are grateful to A. Andreev for drawing our attention to this point}.

To close our discussion we discuss relevant theoretical and experimental works.

Other theoretical work on memory effects has been conducted in the insulating phase. 
In particular, the role of TLS of in memory effects was suggested in Ref.~[\onlinecite{KozubBurin08}], where it was shown that TLS may cause slow relaxation of the local density of states in insulators. The possibility that memory effects can be a manifestation of Anderson Glass\cite{Davies82,Gruenwald82,Thouless77} physics has also been investigated\cite{MerozImry13}. 
Experimentally, memory effects have been found in a variety of systems, including indium oxide films\cite{Ovadyahu91,Ovadyahu93}, thin metallic films\cite{Martinez97}, and granular metals\cite{Grenet03,Grenet07,Grenet10}.
In particular, Ref.~[\onlinecite{Grenet08}] has measured both conductance fluctuations and slow relaxation in samples showing that comparisons of the mesoscopic physics with the memory effets in a single sample are possible. 

We note that the parameter under direct experimental control is the gate voltage, which is related to change in the density by electrostatic considerations which we do not address here.
 However we expect samples with higher density would have increased screening and decreased capacitance. 
This would mean the width of the dip in the conductivity versus gate voltage should be narrower in samples with lower density, in accord with the observation of Ovadyahu\cite{Ovadyahu13}.

\begin{acknowledgments}
 The authors would like to thank O. Agam, A. Andreev,  N. Birge,  Y. Galperin,  L. Glazman, D. Natelson and Z. Ovadyahu for useful comments and suggestions. This work was supported by the Simons foundation.
\end{acknowledgments}

\appendix
\section{The two level systems\label{sec:TLS}}

In this section we give a model for the two level systems. In a disordered system one expects to find a large number of mobile impurities. 
The mobile impurity may be treated as a massive particle which sees a potential $V(r)$ depending on the static impurities and defects in the lattice, as renormalized by electron-phonon excitations.  
We are interested in the case where  $V(r)$ is generally larger than all relevant energy scales, except for localized valleys located an average $r_m$ apart.  
If $r_m$ is large compared to the time scales of our measurement, in a sense to be made precise below, then we expect most of the ``mobile" impurities to not have moved from their valley.  
These are indistinguishable from static impurities.  
However, since the valleys are randomly located we expect to find situations when one impurity sits in a valley, with an unoccupied valley a distance $r\ll r_m$. 
These are the ``close pairs", which are effectively two state systems. We may write down the Hamiltonian for the TLS
\begin{equation}
H_{TLS} = \tilde{\Delta}\sigma_z +  \mathcal{I}\sigma_x,
\label{eq:HCP}
\end{equation}
where $\sigma_{x,y,z}$ are the usual Pauli matrices, and the ``up" states has the impurity localized in one valley, and the ``down" state is the opposite. The level splitting energy $\tilde{\Delta}$ is the difference in the binding energies of the two sites, and $\mathcal{I}$ is the overlap integral. We take $\mathcal{I} = \Lambda_0 e^{-\frac{r}{a}}$ where $\Lambda_0$ is some coupling energy. 

As $\Delta$ and $r$ are properties of the impurities, we take them to be random variables.
Since we are looking for exponentially small terms we may take the random variables to be uniformly distributed without incuring significant error.  We take them to be distributed in the region $\Delta \in  [0,\Delta_m]$, $r\in [0,\ell_{imp}]$. Note we only consider close pairs where $r < \ell_{imp}$ and take this as the upper cutoff on the model. This is taken for convenience so that we may treat all impurities as point scatterers. As longer distances correspond to exponentially longer timescales, there is a well defined regime in which we are insensitive to the details of the cutoff. Since we are only interested in the exponential dependence on $r$ it is sufficient to our accuracy to set $r = \ell_{imp}$ everywhere except in the dependence of $\mathcal{I}$, and we do so in the remainder of this section. 

The close pairs interact with the electrons by altering the local potential. Since this depends on which site the electron occupies, the impurity state and the electronic fluid become coupled. This corresponds to a term in the Hamiltonian 

\begin{equation}
H_{TLS-el} =\frac{\gamma}{2\nu}\left((1+\sigma_z)\psi^{\dagger}_1\psi_1 + (1 - \sigma_z) \psi^{\dagger}_2\psi_2\right).
\label{eq:CPint}
\end{equation}

Here $\gamma$ is the dimensionless interaction strength, $\psi_{1,2}$ is the operator the annihilates a conduction electron at the position $r_{1,2}$, and $r_{1,2}$ are random positions located a distance $r$ apart. We now calculate the time evolution of the density matrix of the close pair, averaging over the metallic system.  This is done most clearly by rotating the sigma matrices so that $H_{TLS}$ is proportional to $\sigma_z$. Working to lowest order in $\mathcal{I}$ this gives: 
\begin{equation}
H_{TLS} = \tilde{\Delta}\tilde{\sigma}_z, \end{equation} 
and, 
\begin{equation}\quad H_{TLS-el} =\left( \tilde{\sigma}_z+\frac{\mathcal{I}}{\tilde{\Delta}}\tilde{\sigma}_x\right)\frac{\gamma}{\nu}\left[\psi^{\dagger}_1\psi_1 - \psi^{\dagger}_2\psi_2\right].
\end{equation}
(plus a sigma independent term). Viewing the electronic fluctuations as a random magnetic field, we see that there is a decohering field and a depolarizing field, where the depolarizing field is smaller by the factor $\mathcal{I}/\tilde{\Delta}$ - exponentially smaller.  Working to second order in the electronic fluctuations we obtain the evolution equation for the density matrix, $\hat{\rho}$. If we parameterize the density matrix by,
\begin{equation}
 \hat{\rho}= \frac{1}{2} + \vec{a}\cdot\vec{\sigma},
\end{equation}
we may give the time evolution by,
\begin{equation}
\frac{\partial \vec{a}}{\partial t} = \Delta \hat{z} \times \vec{a} -\frac{1}{T_2}\vec{a} - \hat{z}\frac{1}{T_1}\left(1 - \text{tanh}(\beta\Delta)\right), \label{eq:CPKinetic}
\end{equation}
 where the energy $\Delta$ is the renormalized level splitting. This depends implicitly on the on the chemical potential, since the compressibilities at $r_1$ and $r_2$ are not equal because of the mesoscopic fluctuations. The decoherence times $T_1$ and $T_2$ are given by
\begin{equation}
T^{-1}_1 = \frac{\gamma^2\mathcal{I}^2}{\Delta^2}\frac{\Delta}{1-\exp(-\Delta/T)} f(\Delta), \label{eq:T1}
\end{equation}
\begin{equation}
T^{-1}_2  = \gamma^2 T f(0)\label{eq:T2},
\end{equation}
where the function $f(\epsilon)$ is $\nu^{-2}$ times the local density-density correlator evaluated at frequency $\epsilon$. This is a function of order unity, with subexponential dependence on $r$. We will therefore treat it as a constant absorbed into $\gamma$.  The dependence on temperature comes from the phase space restricitons on emiting an electron-hole pair, analogous to Korringa\cite{Korringa50} relaxation. 

The behavior of interest happens at time scales much larger then $T_2$, and so the system is effectively classical. Then \niceref{eq:CPKinetic} reduces to a master equation for the diagonal elements of the density matrix $f_{\uparrow} = (1+a_z)/2$ and $f_{\downarrow} = (1-a_z)/2$. The properties of the system will depend on the linear respose functions. Recalling tht the Keldysh function is the autcorrelation and the retarded function is the linear response to change in $\Delta$, we obtain
\begin{equation}
F^K(t) - F^K(0) =\left(\frac{\gamma}{\text{cosh}(\frac{\Delta}{T})}\right)^2\left(1- \exp(-|t|/T_1)\right),\end{equation}
and
\begin{equation}
F^R(t) =\left(\frac{\gamma}{\text{cosh}(\frac{\Delta}{T})}\right)^2\frac{1}{T_1T}\exp(-t/T_1)\Theta(t).\label{eq:CPGreens}
\end{equation}
Again, some smoothly varying function of $r$ has been absorbed into the various constants.  Equation~(\ref{eq:CPGreens}) is in accordance with the classical fluctuation dissipation theorem. 

We will need the ensemble average of the $F$, which we call  $\bar{F} = \ll F \gg$. Let us take the ensemble average over $r$ first, since that contains all of the relevant behavior.  For the Keldysh component,
\begin{equation}\begin{aligned}
\bar{F}^K(t;\Delta) -& \bar{F}^K(0;\Delta)\equiv\left(\frac{\gamma}{\nu\text{cosh}(\frac{\Delta}{T})}\right)^2\\ &\times\,\frac{1}{\ell_{imp} }\int^{\ell_{imp} }_0\! dr\,1-\exp\left[t/t_0\exp(-2r/a)\right],\end{aligned}
\end{equation}
where $t_0$ is a short time scale that depends on $T$ and $\Delta$ from the defintion of $T_1$ in \niceref{eq:T1}. This scale $t_0$ functions as the small time cutoff for the calculations. Changing variables to $\lambda = \exp(-2r/a)$ we obtain,
\begin{equation} \begin{aligned}
\frac{1}{\ell_{imp} }\int^{\ell_{imp}}_{a}\!\!dr&\left\{1-\exp\left[/t_0\exp(-2r/a)\right]\right\}\\ &=\frac{a}{2\ell_{imp} }\int^{1}_{e^{-\frac{2\ell_{imp} }{a}}}\!d\lambda\frac{1-e^{-\lambda t/t_0}}{\lambda}\\
&=\frac{1}{|\log t_m/t_0|}\int^1_{t_0/t_m}\!d\lambda \frac{1-e^{-\lambda t/t_0}}{\lambda}\\ 
&\approx \frac{1}{|\log t_m/t_0|} \int_0^1d\lambda\frac{1-e^{-\lambda t/ t_0}}{\lambda} \\ 
&\approx\frac{\log{t/t_0}}{\log t_m/t_0},
\end{aligned} \end{equation}
where $t_m \equiv t_0\exp(2\ell_{imp }/a)$. The manipulations are valid for times between $t_0$ and $t_m$, which are exponentially seperated. The correlator has a "scale-free" dependence on $t$, which will produce long time correlations. The average of $\Delta$ only smears out the $\log t_m/t_0$ which is insignificant in our regime. The final result is therefore:
\begin{equation}
\bar{F}^K(t)-\bar{F}^K(0) = \frac{\log\left(t/t_0\right)}{\log (t_m/t_0)},\label{eq:AvgFK}
\end{equation}
where we have defined the average scattering time depending on the density of close pairs $\rho^*$,
\begin{equation}
\frac{1}{\tau^*}\equiv\frac{\gamma^2\rho^*}{\nu}\frac{T}{\Delta_m}\tanh(\Delta_m/T).\end{equation}
  The average of $F^R(t)$ can be found simply by taking a time derivative of $\bar{F}^K$
\begin{equation}
\bar{F}^R(t) = \frac{1}{Tt\log(t_m/t_0)}\label{eq:AvgFR}.\end{equation}
The time $\tau^*$ depends linearly on $T$ when $T\ll\Delta_m$. This follows from the fact that only impurities with gaps of order $T$ will be thermally activiated with any probability. This produces the Korringa-like result that $T\tau_*$ is approximately constant at low temperature.

\section{Experimental Protocol\label{sec:Experimental}}
We briefly outline a procedure for detecting the proposed memory effect, in the case of a weak effect in a two dimensional system. We will ignore logarithmic factors throughout this appendix.  

Take a mesoscopic sample of a material with pronounced $1/f$ noise. Measure the scale of the universal conductance fluctuations (UCF), $S_{UCF}$, with magnetic field or gate voltage,
\begin{equation}
S_{UCF} = \langle \left(\frac{\delta I}{I}\right)^2\rangle. \end{equation}
Measure as well the normalized $1/f$ noise, $S_{1/f}$. 
\begin{equation}
S_{1/f}\left(\omega\right) = \frac{1}{I^2} \int dt e^{i\omega(t- t'))} \delta I(t) \delta I(t'). \end{equation}
The strength of the $1/f$ spectrum defines a dimensionless parameter $\alpha$
\begin{equation}
S_{1/f}\left(\omega\right) \sim \alpha\left|\omega\right|^{-1}.
\end{equation}
The ratio of $\alpha$ and the UCF gives the small parameter of our theory,
\begin{equation}
\beta = \alpha/S_{UCF}. \end{equation}
The parameter $\beta$ is approximately the parameter $\left(\frac{1}{T\tau_*}\right)$ that defines the strength of both $1/f$ noise (Eq. 2.18) and the memory effect (Eq. 2.26). 

The memory effect would be obscured by the $1/f$ noise in a mesoscopic sample. To get around this, we use the fact that the  predicted memory does not depend on system size, while the $1/f$ noise decreases like $1/L^2$. So using a large sample of the same material, one could measure the memory dip without the $1/f$ noise.  The predicted depth of the peak in the conductance $\delta G$ is
\begin{equation}
\delta G /G \sim \beta \left(e^2 R_{\Box}/ \hbar\right),
\end{equation} 
\newline
 where  $G$ is the conductance and $R_{\Box}$ is the sheet resistance of the sample.

 There is no upper limit on the size of the sample used to detect the memory dip from the perspective of our mechanism, so the $1/f$ noise may be reduced to arbitrarily, and time averaging can be used to reduce noise on shorter time scales.

\end{document}